\documentclass[aps,prb,twocolumn,groupedaddress]{revtex4}
\usepackage[bookmarks]{hyperref}
\usepackage[dvips]{graphicx}
\usepackage{amsmath}
\usepackage{amstext}
\usepackage{graphics}
\usepackage{exscale}
\usepackage{epsfig}
\usepackage{changebar}
\usepackage[T1]{fontenc}
\usepackage{bm}
\usepackage{bbm}

\usepackage{version}

\begin{document}
\title[Renormalized perturbation theory for n-channel
Anderson model]{Renormalized parameters and perturbation theory for an n-channel
Anderson model with Hund's rule coupling.} 
\author{Y Nishikawa${}^{1,2}$ D J G Crow${}^1$ and A C Hewson${}^1$ }
\affiliation{${}^1$Department of Mathematics, Imperial College, London SW7 2AZ,
  UK.}
\affiliation{${}^2$Graduate School of Science, Osaka City University, Osaka 558-8585, Japan} 

\date{\today}

\begin{abstract}
We extend the renormalized perturbation theory for the single impurity Anderson
model to the $n$-channel model with a Hund's rule coupling, and show
that the exact results for the spin, orbital and charge susceptibilities,
as well as the leading low temperature dependence for the resistivity,
are obtained by working to second order in the renormalized couplings.
A universal relation is obtained between the renormalized parameters,
independent of $n$, in the Kondo regime. 
An expression for the dynamic spin susceptibility is also derived by taking into account
repeated quasiparticle scattering,
 which 
is asymptotically exact in the low frequency regime and satisfies
the Korringa-Shiba relation. The renormalized parameters, including the
renormalized Hund's rule coupling, are deduced from numerical renormalization
group calculations for the model for the case $n=2$. The results
confirm explicitly the universal relations between the parameters
in the Kondo regime. Using these results we evaluate
the   spin, orbital and charge susceptibilities, temperature dependence
of the low temperature resistivity and dynamic spin susceptibility
for the particle-hole symmetric $n=2$ model.
 
\end{abstract}
\maketitle

\section{Introduction}
The single impurity Anderson model~\cite{And61}  has played an important role in
understanding many aspects of the behavior of electrons in systems with
strong electron correlation. Non-perturbative methods have had to  be developed
to make predictions for the behavior of the model in the strong interaction
regime. Among the most successful have been the seminal and pioneering  work of Wilson and
associates \cite{Wil75,KWW80a}  based  on the numerical renormalization group (NRG), and the exact
solutions   using the Bethe ansatz   for the linear dispersion version of the
model \cite{AFL83,TW83}. Though the model was originally put forward to describe magnetic
impurities in a host metal, it has proved to be applicable to other
situations. One main area of application is as a  model for strong correlation 
effects in quantum dots \cite{PG04}. In this application certain  parameters of the model, such as the
 impurity level which determines the
electron occupancy on the quantum dot, can be varied by a gate
voltage.
This  makes it possible  to sweep through different parameter regimes of the
model,
which would be difficult to do for real magnetic impurities, and so the
predictions of the model can be tested more rigorously. The presence of the
narrow many-body resonance in the strong correlation (Kondo)
regime
at low temperatures  can  be  inferred
directly from the measurements of the current through the dot as a function of
an applied bias voltage \cite{GSMAMK98,COK98}. \par
 
Apart from these direct applications of the model, it has also played a role 
in the calculations of strong correlation effects in lattice models. It is
possible to map a class of 
infinite dimensional lattice models of strong electron correlation onto an
effective Anderson impurity model with a self-consistency condition, which 
determines the density of states of the effective medium \cite{GKKR96}. This mapping
requires
that the self-energy is a function of frequency only which is the case  in the 
limit of infinite dimensionality, and the mapping is exact in this limit. For many strongly correlated
systems it is known that the wave vector dependence  of the self-energy is much less 
important than the frequency dependence  so this approach can be used
as a good first approximation for systems in three dimensions (dynamical mean
field theory (DMFT)).  As the  assumption of linear dispersion is not valid
for the effective impurity model generated in
this application, there are no  exact Bethe ansatz
solutions, so the most reliable non-perturbative approaches,
such as the NRG have to be used.\par
It has not proved possible so far to access the strong correlation regime
of the Anderson model by an approach  based purely on perturbation theory in powers of the local
interaction $U$. However, it has been shown that, if the perturbation theory is
reorganized such that the basic parameters of the model are renormalized, then
a perturbation theory in the renormalized interaction $\tilde U$, taken only
to second order gives formally the exact results for the low temperature
properties and low frequency dynamics, provided counter terms are taken into account to avoid over-counting \cite{Hew93,Hew01}. The renormalized parameters have to be determined but these can be
calculated very accurately from an analysis of the low energy excitations 
of an NRG calculation on the approach to the low energy fixed point \cite{HOM04}. 
So far this approach has only been developed in detail for the
non-degenerate one channel model, but the approach is one that
can be applied  to a more general class of models including lattice models.
 Here we extend the calculations to an
$n$-channel impurity Anderson model  with the inclusion of a Hund's rule
exchange term. 
 The Hamiltonian takes the form,  
\begin{eqnarray}
{\cal H}=&&\sum_{m\sigma}\epsilon_{dm\sigma}d^{\dagger}_{m\sigma}d^{}_{m\sigma}+\sum_{k,m\sigma}\epsilon_{km\sigma}
c^{\dagger}_{k m\sigma} c^{}_{k m\sigma}\nonumber \\
&&+\sum_{k m\sigma} (V_k d^{\dagger}_{m\sigma} c^{}_{k m\sigma}
+  V_k^* c^{\dagger}_{k m\sigma} d^{}_{ m\sigma})+{\cal H}_d
\label{model1a}
\end{eqnarray}
where $d^{\dagger}_{ m\sigma}$, $d^{}_{ m\sigma}$, are creation and
annihilation operators for an electron in an impurity state  with total angular momentum
quantum number 
$l$, and $z$-component $m=-l,-l+1,  ... l$, and spin
component
$\sigma=\uparrow,\downarrow$. The impurity level
 in a magnetic field $H$ we take as $\epsilon_{dm\sigma}=\epsilon_d-\mu_{\rm
   B}\sigma H-\mu_{\rm B} mH-\mu$, where  $\sigma=1$ ($\uparrow$) and  $\sigma=-1$ ($\downarrow$) and $\mu$
is the chemical potential, and $\mu_{\rm B}$ the Bohr magneton.  
The creation and annihilation operators $c^{\dagger}_{km\sigma}$, $c^{}_{km\sigma}$ are
for  partial wave conduction electrons with energy
$\epsilon_{km\sigma}$. The  hybridization matrix element for impurity levels with the
conduction electron states is $V_k$. We denote the hybridization width
factor by $\Delta_{m\sigma}(\epsilon)=\pi\sum_k|V_k|^2\delta(\epsilon-\epsilon_{km\sigma})$,
which we can take to be a constant $\Delta$ in the wide flat band limit.
The remaining part of the Hamiltonian,  ${\cal H}_d$ describes the interaction between
the electrons in the impurity state, which we take to be of the form, 
\begin{eqnarray}
{\cal H}_d=&&
    {(U-J_{\rm H})\over 2}\sum_{mm'\sigma\sigma'} d^{\dagger}_{m\sigma}
    d^{\dagger}_{m'\sigma'} d^{}_{ m'\sigma'} d^{}_{ m\sigma}\nonumber  \\
&&+{J_{\rm H}\over 2}\sum_{mm'\sigma \sigma'} d^{\dagger}_{ m\sigma}
d^{\dagger}_{m'\sigma'} d^{}_{m\sigma'} d^{}_{ m'\sigma}. 
\label{model1b}
\end{eqnarray}
As well as the direct Coulomb interaction $U$ between the electrons, we include a Hund's rule
exchange term $J_H$ between electrons in states with different $m$ values.
The sign for the exchange term has been chosen so that $J_H>0$ corresponds
to a ferromagnetic interaction. This model can be used to describe
transition metal impurities, such as Mn or Fe,  in a metallic host in the
absence of spin orbit or crystal field splittings. 
We can interpret the model more generally
with $\alpha=m+l+1$ as a channel index taking values $\alpha=1,2, ...n$ where
$n$ is the number of channels. The Hund's rule term tends to align the
electrons on the impurity site such that for large $U$ and large $J_{\rm H}$
 the impurity state will correspond to a spin $S=n/2$.
The model with $J_{\rm H}=0$  has also been used to describe capacitively
coupled double quantum dots\cite{GLK06}, where the impurity channels correspond to
different
dots.  In that application, however,  the inter-dot interaction
 $U'$ will in general differ from the intra-dot interaction $U$,
so the case here, with $U'=U$, is a special point with
SU(2n) symmetry when $J_{\rm H}=0$.  \par
The structure of this paper will be as follows. In the  next section
we  formulate  the renormalized perturbation theory (RPT) for this model
in terms of the renormalized parameters, $\tilde \epsilon_d$, $\tilde \Delta$, $\tilde U$ and
$\tilde J_H$. We then 
show that  the low temperature behavior,  as measured by the charge and spin
susceptibilities and the low temperature contribution to the resistivity,
can be obtained exactly from the RPT taken to second order in  powers of 
$\tilde U$ and
$\tilde J_H$.
In the localized or Kondo regime we show that   $\tilde \Delta$, $\tilde U$ and
$\tilde J_H$ can be expressed in terms of a single parameter which we
take as the Kondo temperature $T_{\rm K}$. This  relation
is  independent of the channel index 
$n$ and hence applies to all values of $n$. Though we cannot calculate
 $\tilde \Delta$, $\tilde U$ and
$\tilde J_H$  for the general $n$-model using the NRG 
we can calculate them for the two channel case $n=2$. We look at  this case 
in detail and confirm the universal relation
between the renormalized parameters  in the Kondo regime predicted using the 
RPT. 
\section{Renormalized Perturbation Theory}
\begin{figure}[!htbp]
   \begin{center}
     \includegraphics[width=0.4\textwidth]{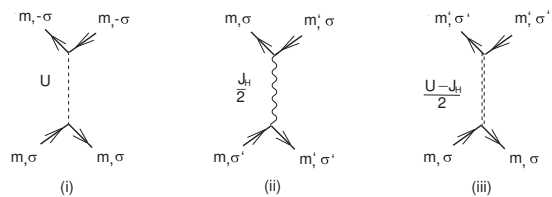}
     \caption{(Color online)  The three interaction vertices corresponding to the terms in the
       Hamiltonian given in Eq. (\ref{model1c}). } 
     \label{vertices}
   \end{center}
 \end{figure}

We start with the Fourier transform of the single particle Green's function
for the impurity $d$-state,
\begin{equation}
G_{d,\sigma}(\omega)=-\int_0^\beta \langle T_\tau
d^{}_{m\sigma}(\tau)d^{\dagger}_{m\sigma}(0)\rangle e^{i\omega_{n'}\tau}\, d\tau,\label{gfdef}
\end{equation}
where $\omega_{n'}=(2n'+1)/\beta$ and $\beta=1/T$ and the brackets $\langle ...
\rangle$ denote a thermal average. 
\begin{equation}
G_{d,m\sigma}(\omega_n)={1\over i\omega_n-\epsilon_{dm\sigma} +i\Delta{\rm sgn}(\omega_n)-\Sigma_{m\sigma}(\omega_n,H)},\label{gf}
\end{equation}
where $\Sigma_{m\sigma}(\omega_n,H)$ is the self-energy. For the zero
temperature Green's function, which will be our main concern, 
$\omega_n$ can be replaced by  continuous variable $\omega$, and summations
over $\omega_n$ replaced by integrations over $\omega$.
  For the perturbation
theory in powers of $U$ and $J_{\rm H}$ it will be convenient to separate the
interaction terms in the Hamiltonian into the terms involving interactions
between electrons in the same channel and those between electrons in
different channels. We rewrite the Hamiltonian from  Eq. (\ref{model1b}) in the form, 
\begin{eqnarray}
{\cal H}_d=&&
    U\sum_{m} n_{d,m\uparrow} n_{d,m\downarrow}\nonumber\\
    &&+ {(U-J_{\rm H})\over 2}\sum_{m\ne m'\sigma\sigma'} d^{\dagger}_{m\sigma}
    d^{\dagger}_{m'\sigma'} d^{}_{ m'\sigma'} d^{}_{ m\sigma}\nonumber  \\
&&+{J_{\rm H}\over 2}\sum_{m\ne m'\sigma \sigma'} d^{\dagger}_{ m\sigma}
d^{\dagger}_{m'\sigma'} d^{}_{m\sigma'} d^{}_{ m'\sigma}. 
\label{model1c}
\end{eqnarray}
The vertices associated with the three types of interaction terms are
illustrated in Fig. \ref{vertices}.\par

 \vspace*{0.7cm}
 \noindent

For the renormalized perturbation theory, the  Green's function in Eq. (\ref{gf}) can be re-expressed as $G_{d,m\sigma}(\omega_n)=z\tilde G_{d,m\sigma}(\omega_n)$,
where $\tilde G_{d,m\sigma}(\omega_n)$ is the quasiparticle Green's function
given by 
\begin{equation}
\tilde G_{d,m\sigma}(\omega_n)={1\over i\omega_n-\tilde\epsilon_{dm\sigma} +i\tilde\Delta{\rm sgn}(\omega_n)-\tilde\Sigma_{m\sigma}(\omega_n,H)}\label{qpgf}
\end{equation}
and the renormalized parameters, $\tilde\epsilon_{dm\sigma}$ and $\tilde\Delta$
are given by
\begin{equation}\tilde\epsilon_{dm\sigma} =z(\epsilon_d+\Sigma_{m\sigma}(0,H))
-\mu_{\rm B}\sigma H-\mu_{\rm B}mH,\quad \tilde\Delta
=z\Delta,\label{rself}
\end{equation}
where $z=1/(1-\partial\Sigma_{m\sigma}(\omega,0)/\partial (i\omega))$
evaluated at $\omega=0$.
The quasiparticle self-energy $\tilde
\Sigma_{m,\sigma}(\omega,H)$ is given by
\begin{eqnarray}&&\tilde\Sigma_{m\sigma}(\omega,H)
  =\nonumber\\&&z\left(\Sigma_{m\sigma}(\omega,H)-\Sigma_{m\sigma}(0,H)\nonumber
-i\omega{\partial\Sigma_{m\sigma}(\omega,0)\over
\partial i\omega}\Big|_{\omega=0}\right ),
\end{eqnarray}
where we have assumed  the Luttinger theorem \cite{Lut60}, ${\rm Im}\Sigma(0)=0$, so that
$\tilde{\rm Im}\Sigma_{m\sigma}(\omega)\sim \omega^2$ as $\omega\to 0$.
When expressed in this form, the $\omega=0$ part of the self-energy and its
derivative have been absorbed into renormalizing the parameters
 $\epsilon_{dm\sigma}$ and $\Delta$,  so in setting up the perturbation 
expansion any further renormalization of these terms must be excluded, or
it will result in over-counting. In working with the fully renormalized
quasiparticles it is appropriate to use
the renormalized  or effective interactions between the quasiparticles.
In the single channel case, we defined the renormalized interaction $\tilde U$
in terms of the four vertex
$\Gamma_{\uparrow,\downarrow,\downarrow,\uparrow}(\omega_1,\omega_2,\omega_3,\omega_4)$ 
in the zero frequency limit \cite{Hew93}.  In this case we need to consider the more
general
four vertex, $\Gamma^{m_1\sigma_1;m_2\sigma_2}_{m_3\sigma_3;m_4\sigma_4}
(\omega_1,\omega_2,\omega_3,\omega_4)$,  which 
 corresponds to the  Fourier coefficient of the connected skeleton diagram for the two particle Green's
 function,
\begin{equation}
\langle T_\tau d^{}_{m_1\sigma_1}(\tau_1)d^{}_{m_2\sigma_2}(\tau_2)d^{\dagger}_{m_3\sigma_3}(\tau_3)
d^{\dagger}_{m_4\sigma_4}(\tau_4)\rangle,
\end{equation}
with the external legs removed. Using the fact that the spin and angular
momentum are conserved independently, and taking into account the antisymmetry conditions of the
fermion creation and annihilation operators, it was shown by Yoshimori
\cite{Yos76} that this vertex at zero frequency  can be expressed in terms of
two parameters,  $\Gamma_C$ and  $\Gamma_e$, as
\begin{eqnarray}
&&\Gamma^{m_1\sigma_1;m_2\sigma_2}_{m_3\sigma_3;m_4\sigma_4}(0,0,0,0)=\nonumber\\
&&\Gamma_C(\delta^{m_1}_{m_4}\delta^{m_2}_{m_3}\delta^{\sigma_1}_{\sigma_4}\delta^{\sigma_2}_{\sigma_3}-\delta^{m_1}_{m_3}\delta^{m_2}_{m_4}\delta^{\sigma_1}_{\sigma_3}\delta^{\sigma_2}_{\sigma_4})\nonumber\\
&&+\Gamma_e(\delta^{m_1}_{m_3}\delta^{m_2}_{m_4}\delta^{\sigma_1}_{\sigma_4}\delta^{\sigma_2}_{\sigma_3}-\delta^{m_1}_{m_4}\delta^{m_2}_{m_3}\delta^{\sigma_1}_{\sigma_3}\delta^{\sigma_2}_{\sigma_4}).
\end{eqnarray}
To first order in the interaction terms, $U$ and $J_{\rm
  H}$,  we have  $\Gamma_C=U-J_{\rm
  H}$ and $\Gamma_e=J_{\rm H}$.
We generalize this result  to specify the renormalized parameters,   $\tilde U$, and $\tilde J_{\rm H}$,
by the relation,
\begin{eqnarray}
&& z^2\,\,
 \Gamma^{m_1\sigma_1;m_2\sigma_2}_{m_3\sigma_3;m_4\sigma_4}(0,0,0,0)=\nonumber\\
&&(\tilde U-\tilde J_{\rm H})(\delta^{m_1}_{m_4}\delta^{m_2}_{m_3}\delta^{\sigma_1}_{\sigma_4}\delta^{\sigma_2}_{\sigma_3}-\delta^{m_1}_{m_3}\delta^{m_2}_{m_4}\delta^{\sigma_1}_{\sigma_3}\delta^{\sigma_2}_{\sigma_4})\nonumber\\
&&+\tilde J_{\rm H}(\delta^{m_1}_{m_3}\delta^{m_2}_{m_4}\delta^{\sigma_1}_{\sigma_4}\delta^{\sigma_2}_{\sigma_3}-\delta^{m_1}_{m_4}\delta^{m_2}_{m_3}\delta^{\sigma_1}_{\sigma_3}\delta^{\sigma_2}_{\sigma_4}),\label{rint}
\end{eqnarray}
 where the factor $z^2$ arises
 from the rescaling of the fields
 to define the quasiparticle Green's function given in Eq.
 (\ref{qpgf}).
For $n=1$ this reduces to 
  \begin{equation}
 z^2\,\, \Gamma^{\sigma_1;\sigma_2}_{\sigma_3;\sigma_4}(0,0,0,0)=\tilde U(\delta^{\sigma_1}_{\sigma_4}\delta^{\sigma_2}_{\sigma_3}-\delta^{\sigma_1}_{\sigma_3}\delta^{\sigma_2}_{\sigma_4}),
\end{equation}
which is the definition of $\tilde U$ used in earlier work \cite{Hew93}.\par

We can combine these terms to define a quasiparticle Hamiltonian $\tilde H$,
\begin{eqnarray}
\tilde H=&&\sum_{m\sigma}\tilde\epsilon_{dm\sigma} \tilde d^{\dagger}_{m\sigma}\tilde d^{}_{m\sigma}+\sum_{km\sigma}\epsilon_{km\sigma}
c^{\dagger}_{km\sigma} c^{}_{km\sigma}\nonumber \\
&&+\sum_{km\sigma} (\tilde V_k \tilde d^{\dagger}_{m\sigma} c^{}_{km\sigma}
+ \tilde V_k^* c^{\dagger}_{km\sigma} \tilde d^{}_{ m\sigma})+\tilde H_d
\label{qpmodel1a}
\end{eqnarray}

\begin{eqnarray}
\tilde H_d=
  &&  {(\tilde U-\tilde J_{\rm H})\over 2}\sum_{mm'\sigma\sigma'} :\tilde d^{\dagger}_{m\sigma}
   \tilde d^{\dagger}_{m'\sigma'} \tilde d^{}_{m'\sigma'} \tilde d^{}_{m\sigma}:\nonumber  \\
&&+\tilde {J_{\rm H}\over 2}\sum_{mm'\sigma\sigma'}: \tilde d^{\dagger}_{m\sigma}
\tilde d^{\dagger}_{m'\sigma'} \tilde d^{}_{m\sigma'} \tilde d^{}_{m'\sigma}:. 
\label{qpmodel1b}
\end{eqnarray}
 The brackets :${\hat O}$: indicate that the operator ${\hat O}$
within the brackets must be normal ordered with respect to the ground state of
the interacting system, which  plays the role of  the vacuum. 
This is because the
interaction terms only come into play when more than one quasiparticle is
created from the vacuum.\par
The renormalized Hamiltonian is not equivalent to the original model, and
the relation between the original and renormalized model is best expressed
in the Lagrangian formulation, where frequency enters explicitly \cite{Hew01}. For
simplicity, we consider the case in the absence of a magnetic field,
where the energy levels $\epsilon_{dm\sigma}$ are independent of $m$ and $\sigma$.
If the Lagrangian density ${\cal L}(\epsilon_d,\Delta,U,J_{\rm H})$ describes the original model,
then by suitably re-arranging the terms we can write
\begin{equation}
{\cal L}(\epsilon_d,\Delta,U,J_{\rm H})={\cal L}(\tilde\epsilon_d,\tilde\Delta,\tilde
U,\tilde J_{\rm H})+{\cal L}_c(\lambda_1,\lambda_2,\lambda_3,\lambda_4),
\end{equation}
where the remainder part ${\cal L}_c(\lambda_1,\lambda_2,\lambda_3,\lambda_4)$
is known as the counter term and takes the form,
\begin{eqnarray}
&&{\cal L}_c(\lambda_1,\lambda_2,\lambda_3,\lambda_4)=
\sum_{m\sigma}\tilde{\bar d}_{m\sigma}(\tau)(\lambda_2\partial_\tau-\lambda_1)
\tilde{ d}_{m\sigma}(\tau)\nonumber\\
 && +(\lambda_3-\lambda_4)\sum_{mm'\sigma\sigma'} \tilde{\bar d}_{m,\sigma}(\tau)
   \tilde {\bar d}_{m'\sigma'}(\tau) \tilde d_{ m'\sigma'}(\tau) \tilde d_{ m\sigma}(\tau)\nonumber  \\
&&+{\lambda_4}\sum_{mm'\sigma \sigma'}\tilde{\bar d}_{ m\sigma}(\tau)
\tilde {\bar d}_{m'\sigma'}(\tau) \tilde d_{m,\sigma'}(\tau) \tilde d_{ m'\sigma}(\tau),
\end{eqnarray}
where $\lambda_1=-\Sigma(0)$, $ \lambda_2=z-1$, $\lambda_3=(z^2U-\tilde U)/2$ and $\lambda_4=(\tilde J_{\rm H}-
z^2 J_{\rm H})/2$.
Though we  can express the coefficients $\lambda_i$, $i=1,2,3,4$, explicitly in terms
of the self-energy terms and vertices at zero frequency, these relations are
not useful in carrying out the expansion. We want to work entirely with the
renormalized parameters and carry out the expansion in powers of
 $\tilde U$ and $\tilde J_{\rm H}$.  We assume that the $\lambda_i$
can be expressed in powers of  $\tilde U$ and $\tilde J_{\rm H}$,
and determine them order by order from the conditions that there should be no further renormalization of quantities taken to be
already 
fully renormalized.  These conditions are
\begin{equation}
\tilde\Sigma_{m\sigma}(0,0)=0,\quad {\partial\tilde\Sigma_{m\sigma}(\omega,0)\over
\partial i\omega}\Big|_0=0,
\end{equation}
and that the renormalized 4-vertex at zero frequency, $\tilde
\Gamma^{m_1\sigma_1;m_2\sigma_2}_{m_3\sigma_3;m_4\sigma_4}(0,0,0,0)$ is such that
\begin{eqnarray}
&&\tilde \Gamma^{m_1\sigma_1;m_2\sigma_2}_{m_3\sigma_3;m_4\sigma_4}(0,0,0,0)=\nonumber\\
&&(\tilde U-\tilde J_{\rm H})(\delta^{m_1}_{m_4}\delta^{m_2}_{m_3}\delta^{\sigma_1}_{\sigma_4}\delta^{\sigma_2}_{\sigma_3}-\delta^{m_1}_{m_3}\delta^{m_2}_{m_4}\delta^{\sigma_1}_{\sigma_3}\delta^{\sigma_2}_{\sigma_4})\nonumber\\
&&+\tilde J_{\rm H}(\delta^{m_1}_{m_3}\delta^{m_2}_{m_4}\delta^{\sigma_1}_{\sigma_4}\delta^{\sigma_2}_{\sigma_3}-\delta^{m_1}_{m_4}\delta^{m_2}_{m_3}\delta^{\sigma_1}_{\sigma_3}\delta^{\sigma_2}_{\sigma_4}).
\end{eqnarray}
In the field theory context these conditions are more commonly known  as the renormalization conditions. They follow directly
from
the definitions of the renormalized self-energy in Eq. (\ref{rself})
and the definitions of the renormalized parameters given  in Eq.
(\ref{rint}).\par
The propagator in the RPT is the free quasiparticle Green's function,
\begin{equation}
\tilde G^{(0)}_{d,m\sigma}(\omega_n)={1\over i\omega_n-\tilde\epsilon_{dm\sigma}
  +i\tilde\Delta{\rm sgn}(\omega_n)}\label{fqpgf}
\end{equation}
The spectral density of the corresponding retarded Green's function gives the free
quasiparticle density of states, $\tilde\rho^{(0)}_{m\sigma}(\omega)$ given by
\begin{equation}
\tilde \rho_{m\sigma}^{(0)}(\omega)={\tilde\Delta/\pi\over (\omega-\tilde\epsilon_{dm\sigma})^2 +\tilde\Delta^2}.\label{fqpdos}
\end{equation}
From Fermi liquid theory,  the quasiparticle interaction terms
 do not contribute to the
linear  specific heat coefficient $\gamma$ of the electrons. It follows that the
 impurity contribution to this coefficient  is proportional to the free quasiparticle
 density of states evaluated at the Fermi level and is given by 
\begin{equation}
\gamma={\pi^2\over 3}\sum_{m,\sigma}\tilde \rho_{m\sigma}^{(0)}(0).\label{shc}
\end{equation}
In the absence of a magnetic field this reduces to $\gamma=2n\pi^2\tilde \rho^{(0)}(0)/3$, where $\tilde \rho^{(0)}(0)$ is the quasiparticle
  density of states per single spin and channel.\par

If we integrate the free quasiparticle density of states in Eq.
(\ref{fqpdos}) to the Fermi level then we get  $ \langle
\tilde n_{dm\sigma}\rangle$ at $T=0$, which is given by 
\begin{equation}
\langle \tilde n_{dm\sigma}\rangle={\eta_{m\sigma}\over \pi}={1\over 2}-{1\over \pi}{\rm tan}^{-1}\left
({\tilde\epsilon_{dm\sigma}\over \tilde\Delta}\right),
\label{qpocc}
\end{equation}
which defines the phase shift $\eta_{m\sigma}$ in the channel with quantum
numbers $m$ and $\sigma$. For this model it has been shown by 
Shiba \cite{Shi75} that $\langle n_{dm\sigma}\rangle=\eta_{m\sigma}/ \pi$,
giving a generalization of the Friedel sum rule, so that
we have $\langle \tilde n_{dm\sigma}\rangle=\langle  n_{dm\sigma}\rangle$;
the quasiparticle occupation number in each channel is equal to the impurity
occupation number in that channel. However, Yoshimori and Zawadowski
\cite{YZ82} have shown that this form of the Friedel sum rule does not hold
for a more general model in which scattering processes can occur between
 $m$-states, $m_1,m_2\to m_3,m_4$, such that   $m_1+m_2=m_3+m_4$. They derive
a restricted form of the sum rule such that $\sum_{m\sigma}a_{m\sigma}
 \langle \tilde n_{dm\sigma}\rangle=\sum_{m\sigma}a_{m\sigma} \eta_{dm\sigma}/\pi$,
where $a_{m\sigma}=1,\sigma,m$. In this more general case, therefore, the
quasiparticle number does not equal the occupation number in the same channel
but we have the more restricted result, $\sum_{m\sigma}a_{m\sigma}\langle 
\tilde n_{dm\sigma}\rangle=\sum_{m\sigma}a_{m\sigma}\langle
n_{dm\sigma}\rangle$.
Using either result, however, we can derive  expressions for the 
zero field spin $\chi_s$, orbital $\chi_{orb}$ and
 charge $\chi_c$ susceptibilities.  We  differentiate  the combinations, $\sum_{m\sigma}a_{m\sigma}
 \langle \tilde n_{d,m\sigma}\rangle$, with $a_{m\sigma}=\sigma, m $ and $1$
respectively, with respect to the magnetic field or in the charge case with
respect to $\tilde\epsilon_d$. To evaluate these expressions we need to calculate
the renormalized self-energy. This calculation taken to first order in
$\tilde U$ and $\tilde J_{\rm H}$ proceeds as in the one channel case
\cite{Hew93,Hew01}, and gives
\begin{equation}
\chi_s=2n\mu_{\rm B}^2\tilde \rho^{(0)}(0)(1+(\tilde U+(n-1)\tilde J_{\rm H})\tilde\rho^{(0)}(0)),
\label{chis}
\end{equation}
\begin{equation}
\chi_{orb}={(n^2-1)\mu_{\rm B}^2\tilde \rho^{(0)}(0)\over 12}\left(1+(\tilde U-3\tilde J_{\rm H})\tilde\rho^{(0)}(0)\right),
\label{chiorb}
\end{equation}
and
\begin{equation}
\chi_c=2n\tilde \rho^{(0)}(0)(1-((2n-1)\tilde U-3(n-1)\tilde J_{\rm H})\tilde\rho^{(0)}(0)).
\label{chic}
\end{equation}

These results can also be obtained from a mean field theory on the 
quasiparticle part of the Hamiltonian given in Eq. (\ref{qpmodel1b})
\cite{Hew93c, Hew93b}. It can be shown using the Ward identities
derived by Yoshimori \cite{Yos76}, which are generalizations of the Ward
identities derived by Yamada \cite{Yam75a,Yam75b} for the single channel case, that
these results are exact. Hence all higher order correction terms in $\tilde U$ and
$\tilde J_{\rm H}$  cancel out. 
\par

In the localized regime a large value of $U$ suppresses the charge
fluctuations on the impurity so $\chi_c\sim 0$. Treating this as an equality,
we get a relation between  $\tilde\rho^{(0)}(0)$, $\tilde U$ and $\tilde J_{\rm H}$,
\begin{equation}
((2n-1)\tilde U-3(n-1)\tilde J_{\rm H})\tilde\rho^{(0)}(0)
=1.
\label{chirelationa}
\end{equation}
When $J_{\rm H}=0$, this reduces to 
\begin{equation}
\tilde U\tilde\rho^{(0)}(0)={1\over (2n-1)},
\label{relationa}
\end{equation}
For the case of half-filling, where $\tilde\epsilon_d=0$ and $\tilde
\rho^{(0)}(0)=1/\pi\tilde\Delta$, the non-linear relation between the
renormalized parameters in Eq. (\ref{chirelationa}) becomes a linear relation between
$\tilde\Delta$, $\tilde U$ and $\tilde J_{\rm H}$,
\begin{equation}
\pi\tilde\Delta=(2n-1)\tilde U-3(n-1)\tilde J_{\rm H}.
\label{chirelation1}
\end{equation}
For $J_{\rm H}=0$ we get
\begin{equation}
\tilde U={\pi\tilde\Delta\over (2n-1)},
\label{relation1}
\end{equation}
which agrees with the one channel result $\tilde U=\pi\tilde\Delta$ for
$n=1$.\par

When  $J_{\rm H}=0$ and we are in the localized limit, we have only one energy scale
which we can take to be the
Kondo temperature, defined for general $n$ such that   $\gamma=\pi^2n/6T_{\rm
  K}$, equivalent to taking  $\tilde\rho^{(0)}(0)=1/4T_{\rm K}$.
 In this limit the result for the
  Wilson ratio, $R_{\rm W}=\pi^2\chi_s/3\mu_{\rm B}^2\gamma=2n/(2n-1)$. This 
 is the same as that for the N-fold degenerate  Anderson model used to describe
 rare earth impurities for  $N=2n$. This result could have
  been anticipated, because the models can be shown to be equivalent
 by putting the orbital $m$ and spin indices
 $\sigma$ into a combined index $\nu=(m,\sigma)$ \cite{GGL09}.
\par
Switching on the interaction  $J_{\rm H}$ ($>0$) will reduce the local orbital
fluctuations,  
as the configuration with the spins aligned will be favored. For $J_{\rm
  H}\gg \pi\Delta$ we can expect the orbital fluctuations to be almost fully
suppressed so that $\chi_{orb}\sim 0$ which, as an equality, gives a further
relation between $\tilde\rho^{(0)}(0)$, $\tilde U$ and $\tilde J_{\rm H}$, 
\begin{equation}
(3\tilde J_{\rm H}-\tilde U)\tilde\rho^{(0)}(0)=1.
\label{chirelationb}
\end{equation}
At half-filling this gives another linear relation between  $\tilde\Delta$, $\tilde U$ and $\tilde J_{\rm H}$, 
\begin{equation}
\pi\tilde\Delta=3\tilde J_{\rm H}-\tilde U.
\label{chirelation2}
\end{equation}
An equivalent condition to that in Eq. (\ref{chirelation2})
can be obtained using the argument
of  Nozi\`eres and
 Blandin \cite{NB80} that the occupation number in a channel
$m$ should be independent of any small change in the chemical potential
in a channel $m'\ne m$ in this regime.
When both the local charge and orbital fluctuations are
suppressed, the
renormalized parameters can be expressed in terms of 
the Kondo temperature $T_{\rm K}$.
 From Eq. (\ref{chirelationa}) and  (\ref{chirelationb}) we deduce
\begin{equation}
\tilde U={3\over 2}\tilde J_{\rm H}=4T_{\rm K}.\label{relationc}
\end{equation}
for the particle-hole symmetric case we have
$1/\tilde\rho^{(0)}(0)=\pi\tilde\Delta=4T_{\rm K}$, so then we have
\begin{equation}
\pi\tilde\Delta=\tilde U={3\over 2}\tilde J_{\rm H}=4T_{\rm K}.\label{relation2}
\end{equation}
which was conjectured  earlier on the basis of a phenomenological mean field
approach \cite{Hew93b,Hew93c}. A notable feature of this result is that 
there is no explicit dependence on $n$. In this regime from Eq.
(\ref{chis})
 we have for the spin susceptibility,
\begin{equation}
\chi_s={(g\mu_{\rm B})^2S(S+1)\over 3T_{\rm K}},
\label{chis2}
\end{equation}
where $S=n/2$ and $g=2$,
which  leads to a Wilson ratio, $R_{\rm W}=2(n+2)/3$ \cite{Yos76,NB80}.\par

\begin{figure}[!htbp]
   \begin{center}
     \includegraphics[width=0.4\textwidth]{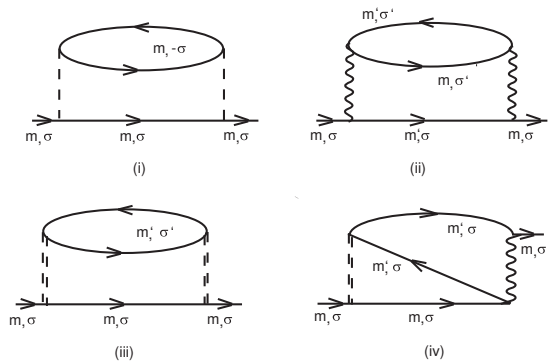}
     \caption{(Color online) Second order diagrams in the renormalized perturbation theory. } 
     \label{2ndorder}
   \end{center}
 \end{figure}

 Yoshimori \cite{Yos76} has  also
derived an exact result for the low temperature 
 impurity contribution to the resistivity in the  particle-hole symmetric
case and  $H=0$. In terms of the renormalized parameters, the result is
\begin{equation}
R(T)=R_0\left(1-{\pi^4(1+2I_R) T^2\over 3}+{\rm O}(T^4)\right),
\label{res1}
\end{equation}
where $I_R$ is given by 
\begin{equation}
I_R= (\tilde \rho^{(0)}(0))^2  ((2n-1)\tilde U^2
-6(n-1)\tilde J_{\rm H}(\tilde U-\tilde J_{\rm H})).
\label{IR}
\end{equation}
This result can be derived in the RPT from a calculation of
the renormalized self-energy $\tilde\Sigma(\omega)$ to second order in $\tilde
U$ and $\tilde J_{\rm H}$.  With the Hund's rule interaction term, there are several types of second order
scattering diagrams which are illustrated in Fig. \ref{2ndorder}. The
vertices are of the same type as shown in Fig. \ref{vertices} but are
weighted by the renormalized interaction terms. 
 The calculations follow along similar lines to those
for  the single channel
case $n=1$ \cite{Hew93,Hew01}. The first order diagrams and the terms linear in $\omega$ are
canceled by the counter terms to this order, and there are no corrections
from the counter terms to the vertices to second order for the case with
particle-hole symmetry. The contributions to
$I_R$ from
diagrams of the types (i) to (iv) respectively in units of $(\tilde \rho^{(0)}(0))^2 $ are:
$\tilde U^2$; $2(n-1)\tilde J^2_{\rm H}$; $2(n-1)(\tilde U-\tilde J_{\rm H})^2$;
-$2(n-1)\tilde J_{\rm H}(\tilde U-\tilde J_{\rm H})$; which give the result in Eq. (\ref{IR}).
 
\par

 \vspace*{0.5cm}
 \noindent

In the localized regime at half-filling the result in Eq. (\ref{res1})
simplifies to give
\begin{equation}
R(T)=R_0\left(1-{\pi^4 (5+4n)\over 96}\left({T\over T_{\rm K}}\right)^2  +{\rm O}(T^4)\right),
\label{res2}
\end{equation}
which agrees with the result derived by Nozi\`eres \cite{Noz74} and Yamada
\cite{Yam75a,Yam75b} for the case $n=1$.
Thus all the exact Fermi liquid relations
can be derived from the RPT  taken to second order only.\par

It was shown in earlier work \cite{Hew06} that the RPT approach can  provide a
description of the dynamic spin susceptibility  for the $n=1$ model   in the low frequency
regime. The calculation  takes account of the repeated
quasiparticle scattering, giving results which are exact in the low frequency
limit $\omega\to 0$, and in remarkably good agreement with the results from a
direct NRG calculation. We extend the calculation to the $n$-channel model
given in Eq. (\ref{model1a}) and (\ref{model1b}). We consider
the Fourier transform of the transverse spin susceptibility,
\begin{equation}
\chi^{+-}_{s}(i\omega_{n'})=\int_0^\beta \langle T_\tau \sum_m S^+_{d,m}(\tau)\sum_{m'}S^-_{d,m'}(0)\rangle e^{i\omega_{n'}\tau}\, d\tau,\label{dynchispin}
\end{equation} 
where $\omega_{n'}=2\pi n'/\beta$ and $S_{d,m}^{+}=
d^{\dagger}_{m\uparrow}d^{}_{m\downarrow}$, $S_{d,m}^{-}=
d^{\dagger}_{m\downarrow}d^{}_{m\uparrow}$ ($S_{d,m}^{z}=(n_{d,\uparrow}-n_{d,\downarrow})/2$).
 We consider the scattering of a
spin up quasiparticle with a spin down quasihole both in channel $m$,
in the absence of a magnetic field. This
particle-hole pair can scatter into a particle-hole pair in the same channel
$m$ or a different channel $m'\ne m$. We consider the scattering into the same
channel first of all. The matrix element for this process is $\tilde U$,
except
we must allow for the fact that $\tilde U$ already takes into account these
processes for $\omega=0$ so, to prevent over-counting, we must use $\tilde
U-\lambda_3$, where $\lambda_3$ is the corresponding counter term. It will
be convenient to use the notation $\bar U$ for $\tilde
U-\lambda_3$. Just taking this type of repeated scattering into account 
gives us a result which has the same form as in the single channel case 
$n=1$ \cite{Hew06},
\begin{equation}
\chi^{+-}_s(\omega+i\delta)={4n\mu_{\rm B}^2}{\tilde\Pi^{+-}(\omega+i\delta)\over 1-
  \bar U\tilde\Pi^{+-}(\omega+i\delta)},\label{chisw}
\end{equation}
where we have analytically continued to real frequency $\omega$. The
free quasiparticle-quasihole propagator in a single 
channel, $\tilde\Pi^{+-}(\omega+i\delta)$, is independent of the channel index
in the absence of a magnetic field, and  is given by
\begin{eqnarray}&&\tilde\Pi^{+-}(\omega+i\delta)={\tilde \Delta\over\pi
      (\tilde\epsilon_d^2+\tilde\Delta^2)},\quad\quad{\omega=0}\nonumber\\
&&={\tilde\Delta\over\pi\omega(\omega+2i\tilde\Delta)}\left\{ {\rm
    ln}\left(1+{\omega \over \tilde\epsilon_d+i\tilde\Delta}\right)\right.\nonumber\\
&&\left.\quad\quad\quad\quad\quad+{\rm ln}\left(1-{\omega \over \tilde\epsilon_d-i\tilde\Delta}\right)\right\}\quad{\omega\ne 0},
\end{eqnarray}
for  $\delta\to +0$. 
We must also take into account that the quasiparticle-quasihole pair being
created in channel $m$  can
scatter into a different channel $m'$, and also be finally annihilated in
a channel with $m'\ne m$. The matrix element for this type of scattering
is $\tilde J_{\rm H}$, corresponding to the diagram in Fig. \ref{vertices} (ii), but again, to avoid over-counting, we replace it
by $\bar J_{\rm H}$. In the absence of  a magnetic field, the
quasiparticle-quasihole propagator is independent of the channel index $m$,
so the summation over the states $m'$ introduces a factor
 $n-1$.   The result of taking these scattering processes into account is that
the 
pair propagator $\tilde\Pi^{+-}(\omega+i\delta)$  in Eq. (\ref{chisw}) is replaced  by
\begin{equation}
{\tilde\Pi^{+-}(\omega+i\delta)\over 1-
  \bar J_{H}(n-1)\tilde\Pi^{+-}(\omega+i\delta)},
\end{equation}
which leads to the result,
\begin{equation}
\chi^{+-}_s(\omega+i\delta)={4n\mu_{\rm B}^2}{\tilde\Pi^{+-}(\omega+i\delta)\over 1-
 (\bar U+(n-1)\bar J_{H})\tilde\Pi^{+-}(\omega+i\delta)}.\label{chiw+-}
\end{equation}
We need to determine the combination $\bar U+(n-1)\bar J_{\rm H}$. We can do
this by requiring that this expression gives $2\chi_s$ in the zero frequency
limit, which is equivalent to the requirement that these scattering
processes contribute to the four vertex at zero frequency are not over-counted. This condition gives
\begin{equation}
\bar U+(n-1)\bar J_{H}={\tilde U+(n-1)\tilde J_{H}\over
1+(\tilde U+(n-1)\tilde J_{H})\tilde \rho^{(0)}(0)}.
\end{equation}
In the Kondo regime this condition simplifies to $\bar U+(n-1)\bar J_{H}=2T_{\rm
  K}(1+2n)/(2+n)$, which gives the one channel result $\bar U=
2T_{\rm
  K}$ for $n=1$.\par

By rewriting Eq. (\ref{chiw+-}) in the form, 
\begin{equation}
{4n\mu_{\rm B}^2\over \chi^{+-}_s(\omega+i\delta) }
={1\over\tilde\Pi^{+-}(\omega+i\delta)} -
 (\bar U+(n-1)\bar J_{H}),
\end{equation}
and taking the imaginary
part, it is straight forward to show that the expression for $\chi_s(\omega)$
satisfies the exact Korringa-Shiba relation,
\begin{equation}
\lim_{\omega\to 0}{{\rm Im}\chi^{+-}(\omega+i\delta)
\over \omega}={\pi\chi^2_s\over n\mu^2_{\rm B}}, 
\label{relation3}
\end{equation}
 which was proved for this model by Shiba \cite{Shi75} and more generally by
Yoshimori and Zawadowsi \cite{YZ82}.\par

So far we have not  discussed how one can calculate the renormalized parameters
 $\tilde\epsilon_d$, $\tilde \Delta$, $\tilde U$ and $\tilde J_{\rm H}$. In the Kondo regime
these reduce to a single parameter $T_{\rm K}$, so one possibility is to  deduce its value
from experiment by fitting the predictions to the measurements  of a  physical
quantity in the low temperature regime, say the impurity susceptibility
or resistivity.  Outside the Kondo regime we have four parameters to determine,
and to calculate all four from experiment one loses much of the predictive
power of the approach. However, it was shown earlier for the single channel
Anderson model how the 
parameters, $\tilde\epsilon_{d}$, $\tilde \Delta$ and $\tilde U$,  can be
calculated in terms of the  bare parameters, $\epsilon_{d}$, $\Delta$ and $U$,
from the many-body low energy excitations 
of an NRG calculation \cite{HOM04}. There are problems in carrying out this procedure 
for the general $n$-channel model, due to the truncation of states which 
has to be carried out in an NRG calculation to reach the very low energy
scales. Truncation means that only  a fraction $1/4^n$ states can be retained at each
NRG iteration. It is possible, however, for the case $n=2$ to compensate for the
lower percentage by increasing the number of states kept at each iteration as the matrices do not get so large. 
In the next section we present for calculations of  $\tilde
\Delta$,
 $\tilde U$, and $\tilde J_{\rm H}$, in terms of  $
\Delta$,
 $U$, and $ J_{\rm H}$, for the $n=2$ model.

\section{NRG Calculation of the Renormalized Parameters for n=2}

 The two-channel model the Hamiltonian ${\cal H}_d$ given in Eq. (\ref{model1b})
can be re-expressed in the form,
\begin{equation}
{\cal H}_d=U\sum_{\alpha=1,2}n_{d\alpha\uparrow}n_{d\alpha\downarrow}
   +U_{12}\sum_{\sigma\sigma'}n_{d,1\sigma}n_{d,2\sigma'}
-{2J_{\rm H}} {\bf S}_{d,1}\cdot{\bf S}_{d,2},  
\label{model2c}
\end{equation}
with a  ferromagnetic Heisenberg exchange coupling $2J_{\rm H}$ between the
electrons in the different channels, and $U_{12}=U-3J_{\rm H}/2$. Our calculations will be restricted to the
particle-hole
symmetric model so we take $\epsilon_d=-U/2-U_{12}$ in the one-electron part of the 
Hamiltonian given in Eq. (\ref{model1a}). The energy of the two
electron triplet state of the isolated impurity with particle-hole symmetry
is $-2U+J_{\rm H}$ and that of the 4-electron or 0-electron state is 0,
so if we are interested in the case when  the triplet state 
is the ground state  configuration, we need to consider the regime $U>J_{\rm H}/2$.
\par
For the NRG calculations the model is recast in a form such that the
impurity is
coupled via a hybridization $V$ to two tight binding
chains which describe the conduction electron states, one chain for
each channel. The conduction electron band is discretized
with a discretization parameter $\Lambda>1$, such that the 
couplings decrease along the chains as $\Lambda^{-N/2}$ for large $N$,
where $N$ is the $N$th site along the chain from the impurity.
The calculations are then carried out iteratively by direct diagonalization,
starting at the impurity site and  adding one further site to each chain
at each iteration step. The number of basis states used has to be truncated
when the matrices get too large for  diagonalization on a practical timescale, which can occur
after only a few iterative steps. When truncation is applied a fixed number
 of states is retained at each step.
For the $n=2$ model considered here, we take 3600 states, which is a
factor of 3 to 4 more than for the non-degenerate model ($n=1$) and
a discretization factor $\Lambda=6$.
 We can check the expected accuracy of our calculations by 
using this value for $\Lambda$ to calculate $\tilde U$ and $\pi\tilde\Delta$ for the single channel
model and compare with the values deduced indirectly from the exact Bethe
ansatz results for the specific heat coefficient $\gamma$ and the zero
temperature spin
susceptibility \cite{HOM04}. For $U/\pi\Delta=2$, $\pi\Delta=0.01$, keeping $900$
states, we get the values,  $\tilde U=0.2295$ and $\pi\tilde\Delta=0.2387$, which
can be compared
 with those deduced from the  Bethe ansatz,   $\tilde
U=0.2301$ and $\pi\tilde\Delta=0.2392$. This gives an accuracy of better
than  0.3\%.
For further details on setting up the NRG calculations, we refer to the original papers  \cite{Wil75,KWW80a}
and the  recent  review article \cite{BCP08}.\par
With this discrete spectrum the Green's function in Eq. (\ref{gf})
takes the form,
\begin{equation}
G_{d,\sigma}(\omega)={1\over i\omega-\epsilon_{dm\sigma}-|V|^2g_{\alpha\sigma}(i\omega)-\Sigma_{m\sigma}(\omega)},\label{nrggf}
\end{equation}
where $g_{\alpha\sigma}(i\omega)$ is the Green's function for the
first site for the isolated conduction band chain.\par
The connection between the NRG approach and the renormalized perturbation
theory is based on identifying the quasiparticle Hamiltonian,
given in Eq. (\ref{qpmodel1a}) and (\ref{qpmodel1b}), as the low energy fixed point of the NRG together with the leading
irrelevant terms \cite{Hew05}. The lowest single-particle excitations from
the NRG ground state  should  correspond to a quasiparticle
excitation described by the one-body part of the quasiparticle Hamiltonian
as given in Eq. (\ref{qpmodel1a}). For the calculation of the interaction
terms, $\tilde U$ and $\tilde J_{\rm H}$, from the NRG we have to consider
the 
difference between two-body excitations from the 
NRG ground state and the two corresponding one-body excitations.
\par
  
The low energy  single-particle excitations are given by the
poles of the non-interacting quasiparticle Green's function when analytically
continued to real frequency $\omega$.
The equation for these poles is the same as  that for the non-interacting 
model but with a renormalized hybridization $\tilde V$ and
energy level $\tilde\epsilon_d$. Therefore, the lowest energy single
particle and hole excitations, $E_p(N)$ and $E_h(N)$, from the {\it interacting }
ground state  should be solutions of the equation,
 \begin{equation}
 \omega-\tilde \epsilon_{d}
  -|\tilde V|^2g_{\alpha\sigma}(\omega)=0.\label{qpe}
\end{equation}
If we substitute the excitations energies, $E_p(N)$ and $E_h(N)$,
as calculated in the NRG for a finite 
chain length $N$, into Eq. (\ref{qpe}) then we can deduce corresponding $N$-dependent
renormalized parameters $\tilde V(N)$ and $\tilde \epsilon_d(N)$.
Only if  $\tilde V(N)$ and $\tilde \epsilon_d(N)$
become independent of $N$ for large $N$,
do the low energy one-particle energy levels of the interacting
system correspond to those of a renormalized non-interacting model.
 If this is the case, then
the asymptotic values for large $N$ define the renormalized
parameters $\tilde V$ (and hence $\tilde \Delta$) and
$\tilde \epsilon_d$.\par
To calculate the renormalized interaction terms, we first have to
diagonalize the non-interacting  impurity model with the renormalized
parameters, which describes the quasiparticles. The interaction terms
are then added to the quasiparticle Hamiltonian and expressed
using the diagonalized single quasiparticle states as a basis.
The energy difference between the lowest two-particle state
and the sum of the corresponding two quasiparticle states
is equal to the expectation value of interaction terms
in the quasiparticle Hamiltonian. The interaction parameter
$\tilde U$ can be calculated from the NRG results for the lowest two-particle
excitation in the {\it same} channel which will be independent of $\tilde J_{\rm H}$. 
For a finite length chain
$N$ the value $\tilde U(N)$ will depend upon $N$, and
for this to correspond to a low energy quasiparticle Hamiltonian
$\tilde U(N)$ should become independent of $N$ for large $N$.
The asymptotic values of $\tilde U(N)$ for large $N$ defines
the renormalized parameter $\tilde U$.  Similarly, to calculate
$\tilde J_{\rm H}$ we  look at the difference between the single and
triplet states of a two-particle excitation with one electron
excitation in each of the two channels. This excitation 
 will be independent of $\tilde U$ and depend only on $\tilde J_{\rm H}$.
Using the NRG results for a finite chain of  $N$ sites,
 we can define a parameter $\tilde J_{\rm H}(N)$, with $\tilde J_{\rm H}$ 
 given by the asymptotic value of $\tilde J_{\rm H}(N)$ for large $N$.
Further details on the calculations of the renormalized parameters from the low
energy NRG states can be found in reference \cite{HOM04}.\par

\begin{figure}[!htbp]
   \begin{center}
     \includegraphics[width=0.49\textwidth]{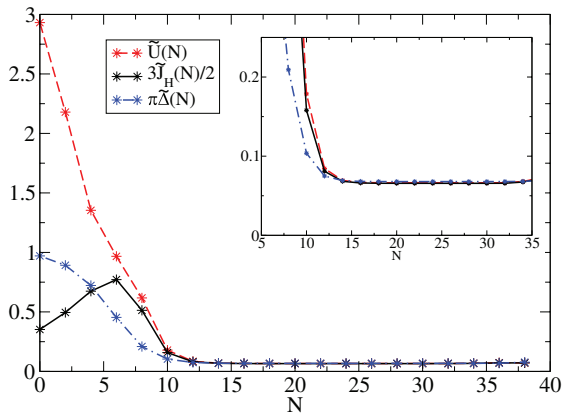}
     \caption{(Color online) A plot of $\tilde\Delta(N)/\pi\Delta$, $\tilde U(N)/\pi\Delta$
and  $3\tilde J_{\rm H}(N)/2\pi\Delta$
  versus $N$ for  $ U/\pi\Delta=3.6$
$J_{\rm H}/\pi\Delta=0.15$ and  $\pi\Delta=0.01$. The inset shows the convergence of
  these parameter to a common limit in this case as the bare parameters
 correspond to a point in the Kondo regime. } 
     \label{fig1}
   \end{center}
 \end{figure}

We first show results for the renormalized parameters as a function of
$N$. We show a typical case in Fig. \ref{fig1} for the parameters
 $\tilde U(N)$, $\pi\tilde\Delta(N)$ and $3\tilde
J_{\rm H}(N)/2$ as a function of $N$ for  $\pi\Delta=0.01$, $U/\pi\Delta=3.6$
and $J_{\rm H}/\pi\Delta=0.15$, which is a parameter set corresponding to a
point in the Kondo regime where the orbital fluctuations have been suppressed. 
The results demonstrate that not only is there
 a plateau region for all the parameters for large $N$, but also that the asymptotic values
of  $\tilde U(N)$, $\pi\tilde\Delta(N)$ and $3\tilde
J_{\rm H}(N)/2$ correspond to a single energy scale and satisfy the relation
given in Eq. (\ref{relation2}). The choice of a relatively large value of
$\Lambda=6$ means that the convergence to a plateau region is achieved for
relatively
small values of $N$. The plateau region is finite because the renormalized parameters
correspond to the leading irrelevant corrections to the free fermion fixed point
of the Wilson renormalization group transformation \cite{Wil75}, so eventually they diverge
from the plateau when $N$  is such that the decreasing irrelevant corrections
become of the same order as the uncertainties in the numerical  computation.

\par
\vspace*{0.7cm}
 
\begin{figure}[!htbp]
   \begin{center}
     \includegraphics[width=0.47\textwidth]{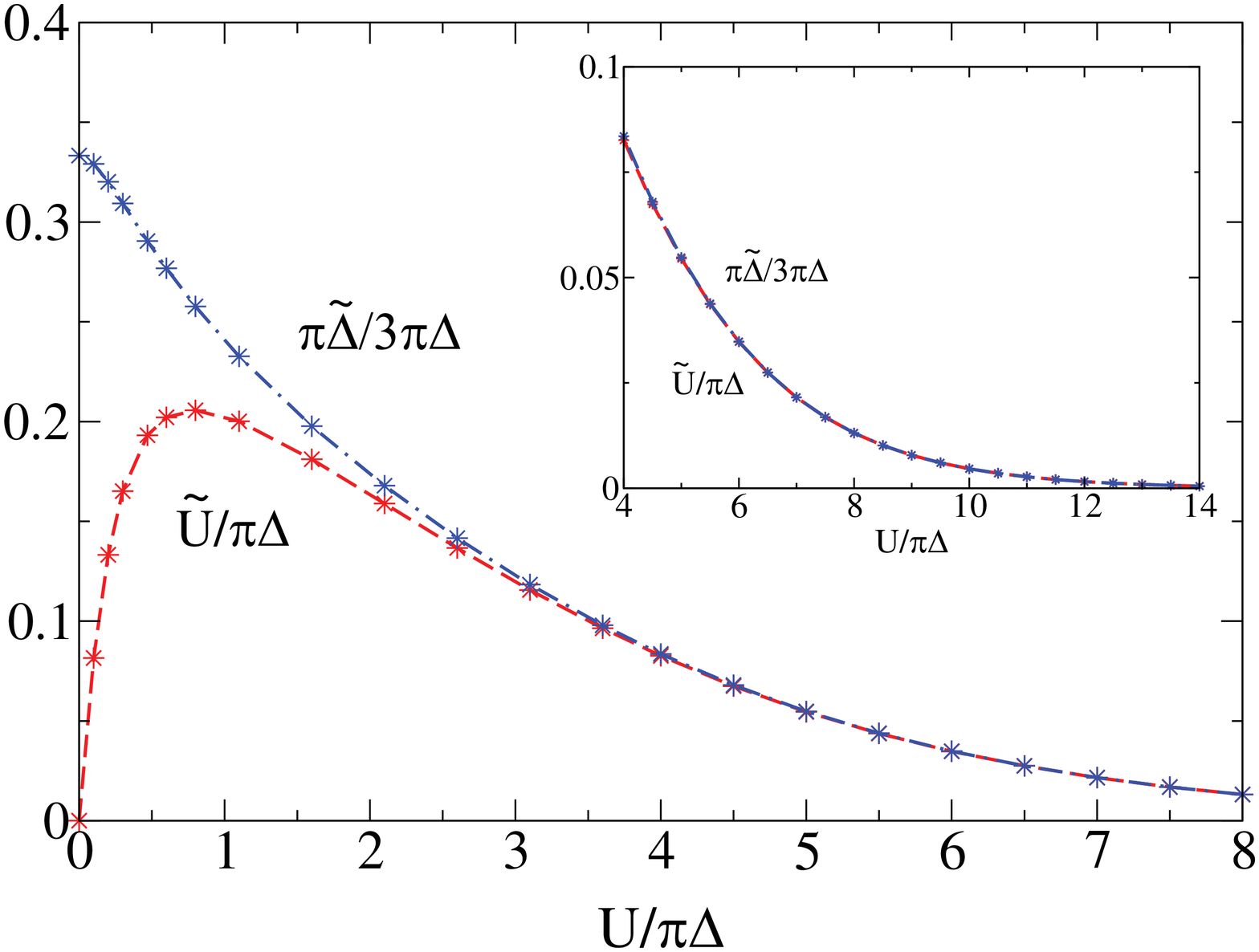}
     \caption{(Color online) A plot of $\tilde\Delta/3\pi\Delta$ and  $\tilde U/\pi\Delta$
  versus $U/\pi\Delta$ for $J_{\rm H}=0$ and  $\pi\Delta=0.01$. } 
     \label{fig2}
   \end{center}
 \end{figure}

We next look at the renormalized parameters in the different parameter
regimes of the model. In Fig. \ref{fig2} we show the results for
$\tilde\Delta/3\Delta$ and  $\tilde U/\pi\Delta$
  versus $U/\pi\Delta$ for $J_{\rm H}=0$  ($\pi\Delta=0.01$).
We predicted from Eq. (\ref{relation1}) that for large $U/\pi\Delta$
 we should have a single energy scale such that for $n=2$,
 $\tilde U=\pi\tilde\Delta/3$ and the results clearly show that this is the case
 for $U/\pi\Delta>3$.  The numerical results for the ratio  $\tilde
 U/\pi\tilde\Delta$ for large $U$
give the value $1/3$ to an  accuracy of 0.01\%.\par
\begin{figure}[!htbp]
   \begin{center}
     \includegraphics[width=0.47\textwidth]{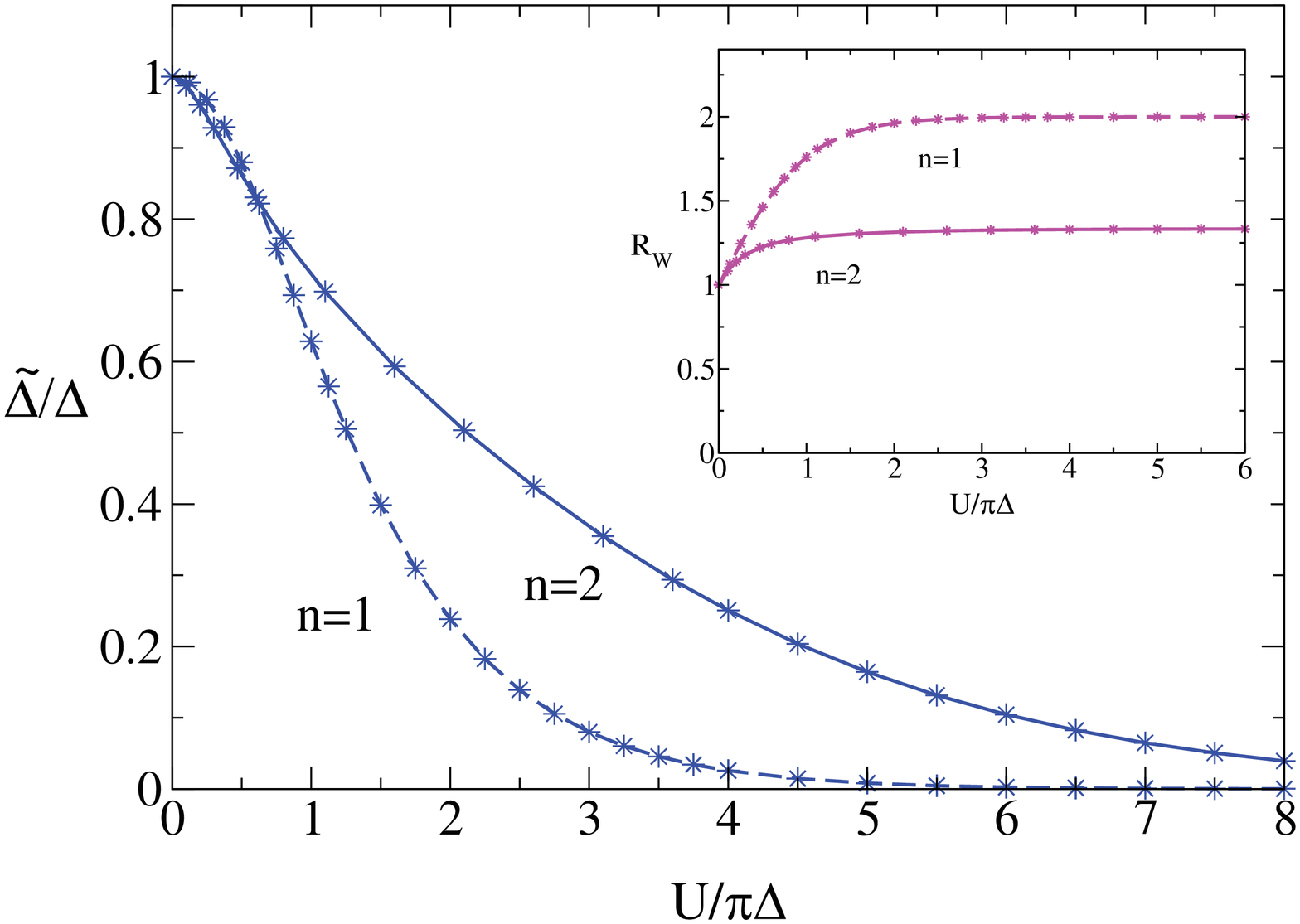}
     \caption{(Color online) A comparison of $\tilde\Delta/\Delta$ 
  versus $U/\pi\Delta$ for the $n=1$ and $n=2$ models for  $J_{\rm H}=0$ and
  $\pi\Delta=0.01$. The inset shows the corresponding values for the Wilson
ratio $R_{\rm W}=1+\tilde U/\pi\tilde\Delta$.} 
     \label{fig2.5}
   \end{center}
 \end{figure}
\begin{figure}[!htbp]
   \begin{center}
     \includegraphics[width=0.42\textwidth]{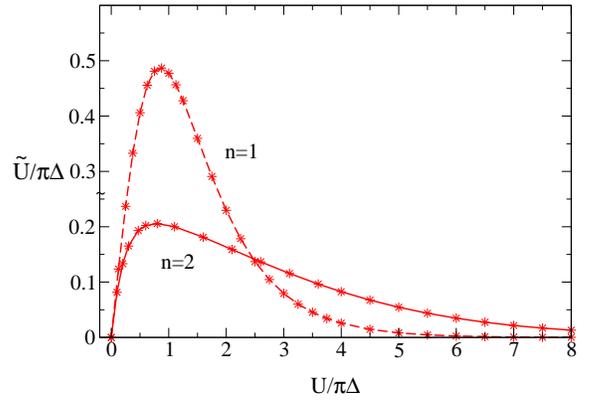}
     \caption{(Color online) A comparison of $\tilde U/\pi\Delta$ 
  versus $U/\pi\Delta$ for the $n=1$ and $n=2$ models for  $J_{\rm H}=0$ and  $\pi\Delta=0.01$} 
     \label{fig2.6}
   \end{center}
 \end{figure}

 In Fig. \ref{fig2.5} and \ref{fig2.6} we compare the results
for these two quantities with those for the single channel model $n=1$. 
We can see that the parameters $\tilde\Delta$ and  $\tilde U$
 the fall off
with increase of $U$  much more slowly for the two channel model.
 This is because
in the two channel model we have unsuppressed fluctuations of the orbital
component. When $J_{\rm H}=0$ and finite $U$, at half-filling in the isolated
impurity for the two channel model there are six degenerate two-electron configurations
with energy $2\epsilon_d+U$. 
 Both the $n=1$ and $n=2$ models
in the Kondo regime can be described by localized  SU(2n) Kondo model. For the case
$n=1$ it is the s-d or SU(2) Kondo model and for $n=2$ the Coqblin-Schrieffer or SU(4) Kondo model. The Hamiltonian for the SU(2n) Kondo model takes
the form,
\begin{equation}
{\cal H}_K(2n)= J_{\rm eff} \sum_{\nu,\nu',k,k'}Y_{\nu,\nu'}c^{\dagger}_{k',\nu'}c^{}_{k,\nu}
+ \sum_{\nu,k}\epsilon_{k}c^{\dagger}_{k,\nu}c^{}_{k,\nu},
\label{Kondo_model}\end{equation}
where the sum over $\nu=1,2, ... 2n$, and  with particle -hole symmetry
 $J_{\rm eff}=
4|V|^2/U$. The operators $Y_{\nu,\nu'}$ obey
the SU(2n) commutation relations,
\begin{equation}
  [Y_{\nu,\nu'}, Y_{\nu'',\nu'''}]_-=Y_{\nu,\nu'''}\delta_{\nu',\nu''}-Y_{\nu'',\nu'}\delta_{\nu,\nu'''},
\end{equation}
with  $\sum_\nu Y_{\nu,\nu}=nI$. For $n=1$, $Y_{\nu,\nu'}=|\nu\rangle\langle \nu'|$, where
$|\nu\rangle$
are the single electron impurity states with spin up ($\nu=1$) and spin down
($\nu=2$), giving a two dimensional representation for the $Y_{\nu,\nu'}$.
In the two channel case for half-filling the representation of the operators
 $Y_{\nu,\nu'}$ is six dimensional and details of the $Y_{\nu,\nu'}$ in terms of the
two electron impurity states are given in the Appendix.\par

\begin{figure}[!htbp]
   \begin{center}
     \includegraphics[width=0.42\textwidth]{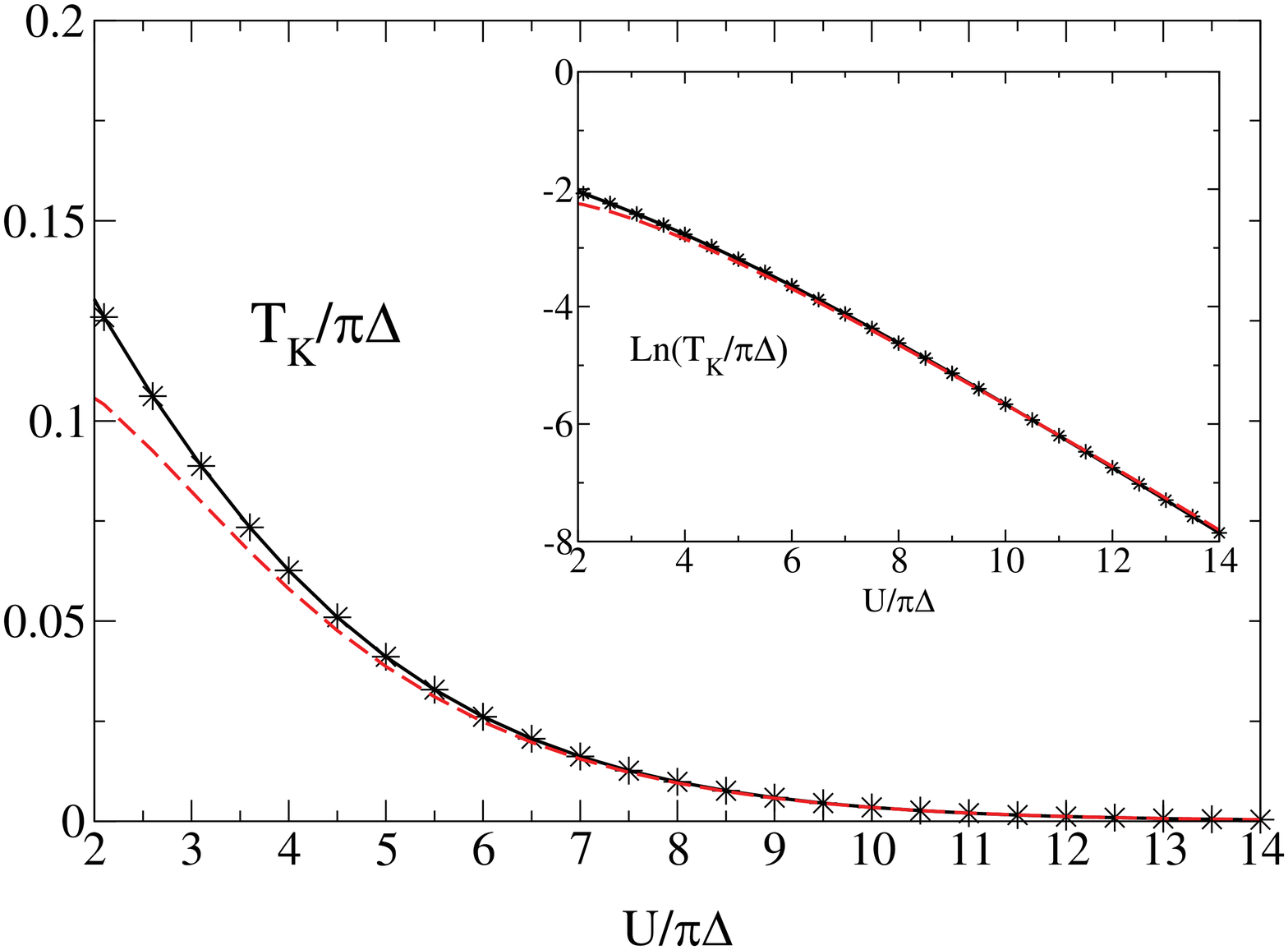}
     \caption{(Color online) A plot of $T_{\rm K}/\pi\Delta$ ($\pi\tilde\Delta=4T_{\rm K}$)
as a function of $U/\pi\Delta$
for $J_{\rm H}=0$ and $\pi\Delta=0.01$. The dashed curve corresponds to the
formula ${1.01u}e^{-\pi^2u/
  16+0.25/u}/2\pi$, where $u=U/\pi\Delta$. The inset shows a plot of the
  logarithm for the same two curves.
} 
     \label{fignutk}
   \end{center}
 \end{figure}

The relation   $T_{\rm K}=\pi\tilde\Delta/4$ applies for both the $n=1$
and $n=2$ models in the case of particle-hole symmetry. 
 In the single
channel case $n=1$,  $T_{\rm K}$ is known from the Bethe ansatz solution,
and is given by $T_{\rm K}/\pi\Delta=
\sqrt{u/2\pi}e^{-\pi^2u/8+0.5/u}$, where $u=U/\pi\Delta$ \cite{HZ85}, and the NRG
results for  $T_{\rm K}$  deduced from $\pi\tilde\Delta$ are in precise agreement with this expression for large $U$.
 For $n\ge 2$ there is no Bethe ansatz solution for the model with finite $U$.
However, there is  a Bethe ansatz solution for the SU(N) Kondo model
(Coqblin-Schrieffer model) and the N-fold degenerate
Anderson model with $U=\infty$ \cite{AD84,TW84,OTW83}
  which gives in  the exponential for $T_{\rm
  K}$ a factor  proportional to
 $1/N$. The prefactor is not universal and depends on the cut-offs 
 used for the high energy excitations in the model.
 We have taken for the
two channel case, therefore, 
the expression $T_{\rm K}/\pi\Delta=1.01 ue^{-\pi^2u/16+0.25/u}/2\pi$,
where the prefactor has been chosen to give the most reasonable fit to the
data. The result of this fitting is shown in Fig. \ref{fignutk}, where it can
been seen that the agreement is very good in the strong coupling range 
$U/\pi\Delta>4.0$. The same form for   $T_{\rm K}$ was used in
reference \cite{GGL09}, and found to be in good agreement with their NRG results.
 \par

\vspace*{0.7cm}

\vspace*{0.9cm}
 \noindent

\vspace*{1.9cm}
\begin{figure}[!htbp]
   \begin{center}
     \includegraphics[width=0.45\textwidth]{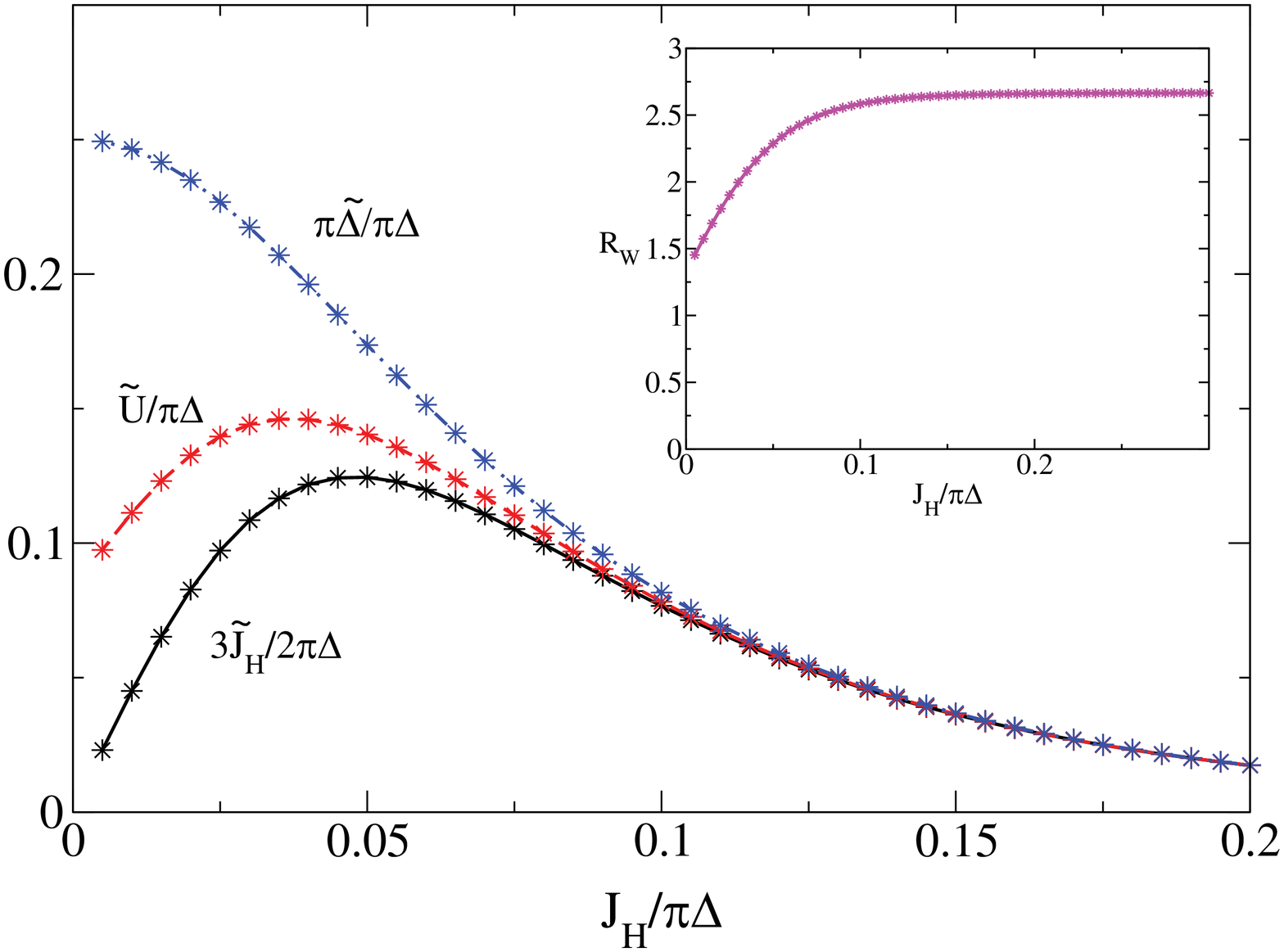}
     \caption{(Color online) A plot of $\pi\tilde\Delta/\pi\Delta$, $\tilde U/\pi\Delta$
 $3\tilde J_{\rm H}/2\pi\Delta$ versus $J_{\rm H}/\pi\Delta$ for
 $U/\pi\Delta=4.0$ and $\Delta=0.01$. There is a single renormalized energy
scale when $J_{\rm H}/\pi\Delta>0.1$. The inset shows the corresponding Wilson
ratio $R_{\rm W}=1+(\tilde U+\tilde J_{\rm H})/\pi\tilde\Delta$.
} 
     \label{fig3}
   \end{center}
 \end{figure}

 \noindent
In Fig. \ref{fig3} we look at the effect of switching on the Hund's rule
term $J_{\rm H}$ for a relatively large value of $U$, $U/\pi\Delta=4.0$, which
is sufficient to suppress the charge fluctuations. As we increase $J_{\rm H}$,
we begin to suppress also the orbital fluctuations, such that when $J_{\rm
  H}/\pi\Delta>0.1$
we are in the regime where we have a single energy scale. This we refer to as
the Kondo regime with the Kondo temperature in the particle-hole symmetric
case  given by $\pi\tilde\Delta=4T_{\rm
  K}$. In this regime the relations between the renormalized parameters  are
such that  $\tilde U_{12}=\tilde U -3\tilde J_{\rm H}/2=0$.\par
\vspace*{0.9cm}
 \noindent

\begin{figure}[!htbp]
   \begin{center}
     \includegraphics[width=0.41\textwidth]{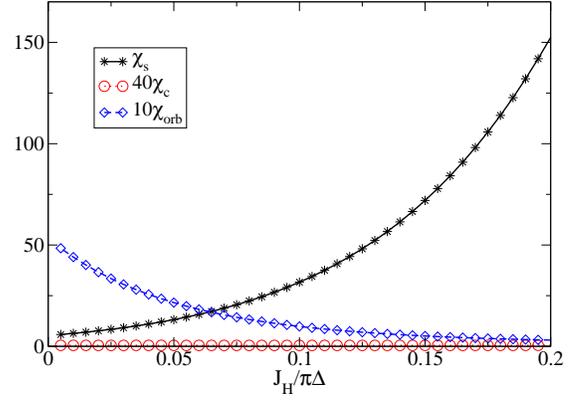}
     \caption{(Color online) A plot of  the spin susceptibility $\chi_s$ (units of $4\mu_{\rm B}^2$),
       $40\times \chi_c$, where $\chi_c$ is the charge
       susceptibility, and the  $10\times \chi_{orb}$, where $\chi_{orb}$ is the orbital
       susceptibility  (units of $\mu_{\rm B}^2/4$),
 versus $J_{\rm H}/\pi\Delta$ for the same parameter set as in Fig. \ref{fig3}.
} 
     \label{fig4}
   \end{center}
 \end{figure}

In Fig.  \ref{fig4} we plot the corresponding spin, orbital and
charge susceptibilities using the expression for these given in Eq.
(\ref{chis}), (\ref{chiorb}) and (\ref{chic}) for the set of parameters used
for Fig. \ref{fig3}.
The fact that the charge susceptibility is almost zero, due to the large value
of $U$, $U/\pi\Delta=4.0$, means that the renormalized parameters must satisfy Eq. (\ref{chirelation1}).
This provides some insight into why the value of $\tilde U$ increases initially
as $J_{\rm H}$ is switched on. For small $J_{\rm H}$ the change in  $\tilde
J_{\rm H}$
is almost linear whereas the change in $\pi\tilde\Delta$ is relatively small.
Therefore to satisfy Eq. (\ref{chirelation1}) $\tilde U$ must also increase almost
linearly
in this region. We can also see from Fig. \ref{fig4} that the orbital
susceptibility
is small (multiplied by a factor 10 in the figure), and decreases
monotonically
as $J_{\rm H}$ increases.\par
\vspace*{0.7cm}
\begin{figure}[!htbp]
   \begin{center}
     \includegraphics[width=0.47\textwidth]{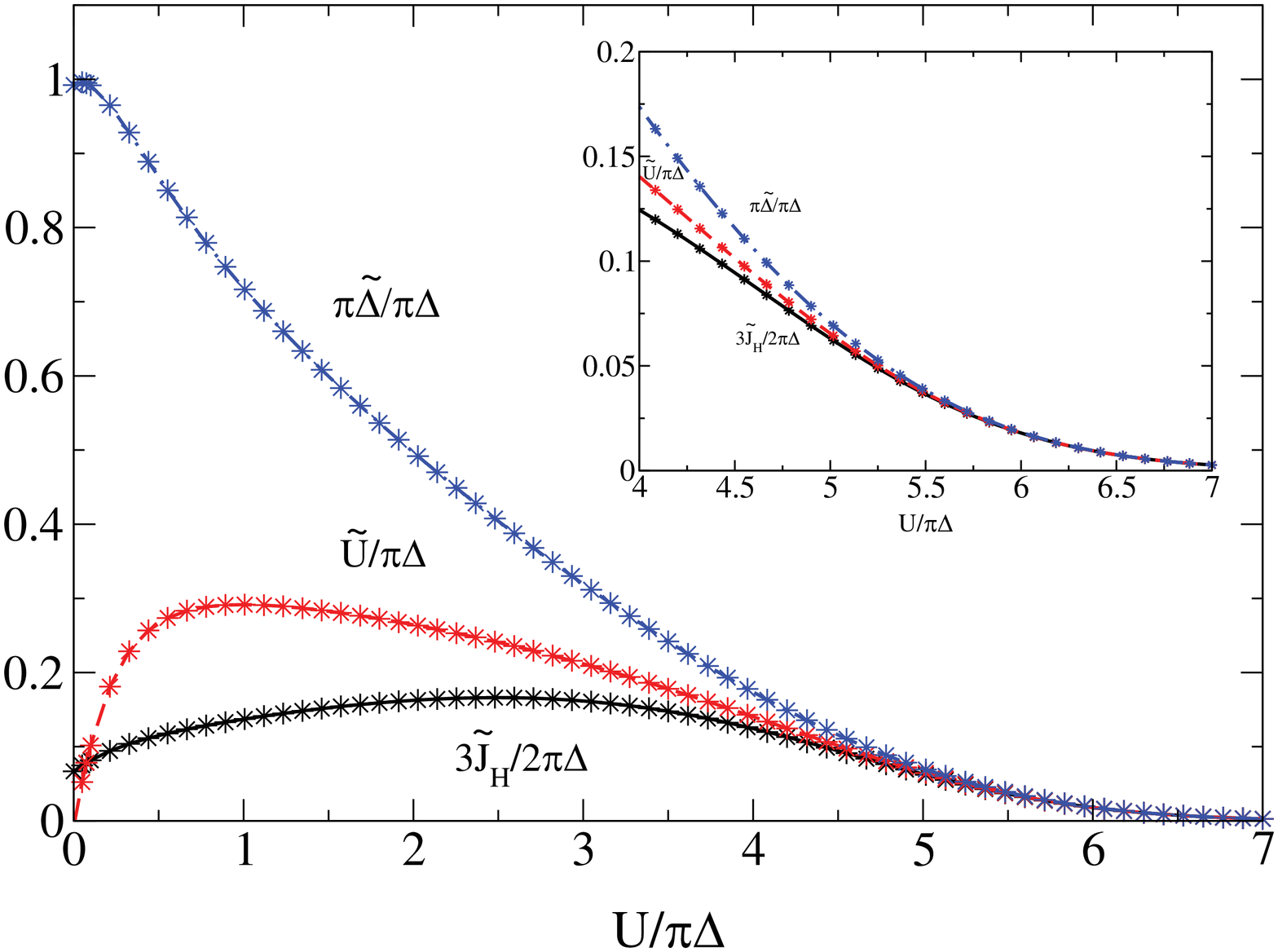}
     \caption{(Color online) A plot of $\pi\tilde\Delta/\pi\Delta$, $\tilde U/\pi\Delta$
and  $3\tilde J_{\rm H}/2\pi\Delta$ versus $U/\pi\Delta$ 
for $J_{\rm H}/\pi\Delta=0.05$ and $\Delta=0.01$. } 
     \label{fig5}
   \end{center}
 \end{figure}

\begin{figure}[!htbp]
   \begin{center}
     \includegraphics[width=0.36\textwidth]{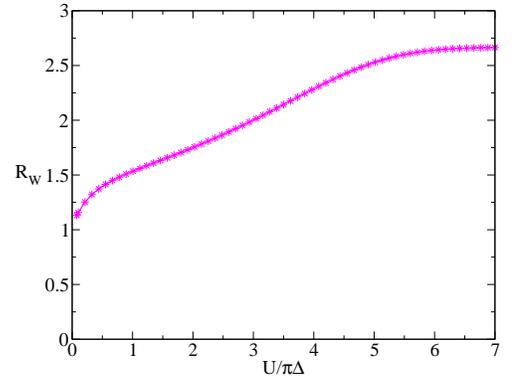}
     \caption{(Color online) A plot of the Wilson ratio $R_{\rm W}=1+(\tilde U+\tilde J_{\rm H})/\pi\tilde\Delta$ versus $U/\pi\Delta$ 
for $J_{\rm H}/\pi\Delta=0.05$ and $\Delta=0.01$.} 
     \label{fig5a}
   \end{center}
 \end{figure}

 \noindent
In Fig. \ref{fig5} we explore
a different parameter regime. Here the parameters
 $\tilde\Delta$, $\tilde U$, and 
 $3\tilde J_{\rm H}/2$ are plotted for a range of values of $U$
for $J_{\rm H}/\pi\Delta=0.05$. We see that for this smaller value of $J_{\rm
  H}$ a large value of $U/\pi\Delta\sim 5.5$ is required before the orbital
fluctuations are suppressed and the Kondo regime is achieved. We suggest that
the explanation for this behavior is that a large $U$ strongly renormalizes the effective hopping
parameter, which is proportional to $\sqrt{\tilde \Delta}$, so that the relatively weaker
$J_{\rm H}$ is then sufficient to suppress the orbital fluctuations. The inset
of Fig. \ref{fig5} shows that the renormalized parameters  do actually converge for
$U/\pi\Delta>5.5$ to a common value. In Fig. \ref{fig5a} we plot the Wilson
$\chi_s/\gamma$
ratio, $R_{\rm W}=1+(\tilde U+\tilde J_{\rm H})/\pi\tilde\Delta$, for the
parameter set given in Fig. \ref{fig5}. It shows a steady increase from a
value $R_{\rm W}\sim 1$ for small $U$ with a leveling off at $U/\pi\Delta\sim
5$ and then a convergence to the value $R_{\rm W}=8/3$, corresponding to 
that of the localized $S=1$  two channel Kondo model.
\par

\vspace*{0.7cm}
 \noindent

\vspace*{0.7cm}
 \noindent

\vspace*{0.7cm}

\begin{figure}[!htbp]
   \begin{center}
     \includegraphics[width=0.42\textwidth]{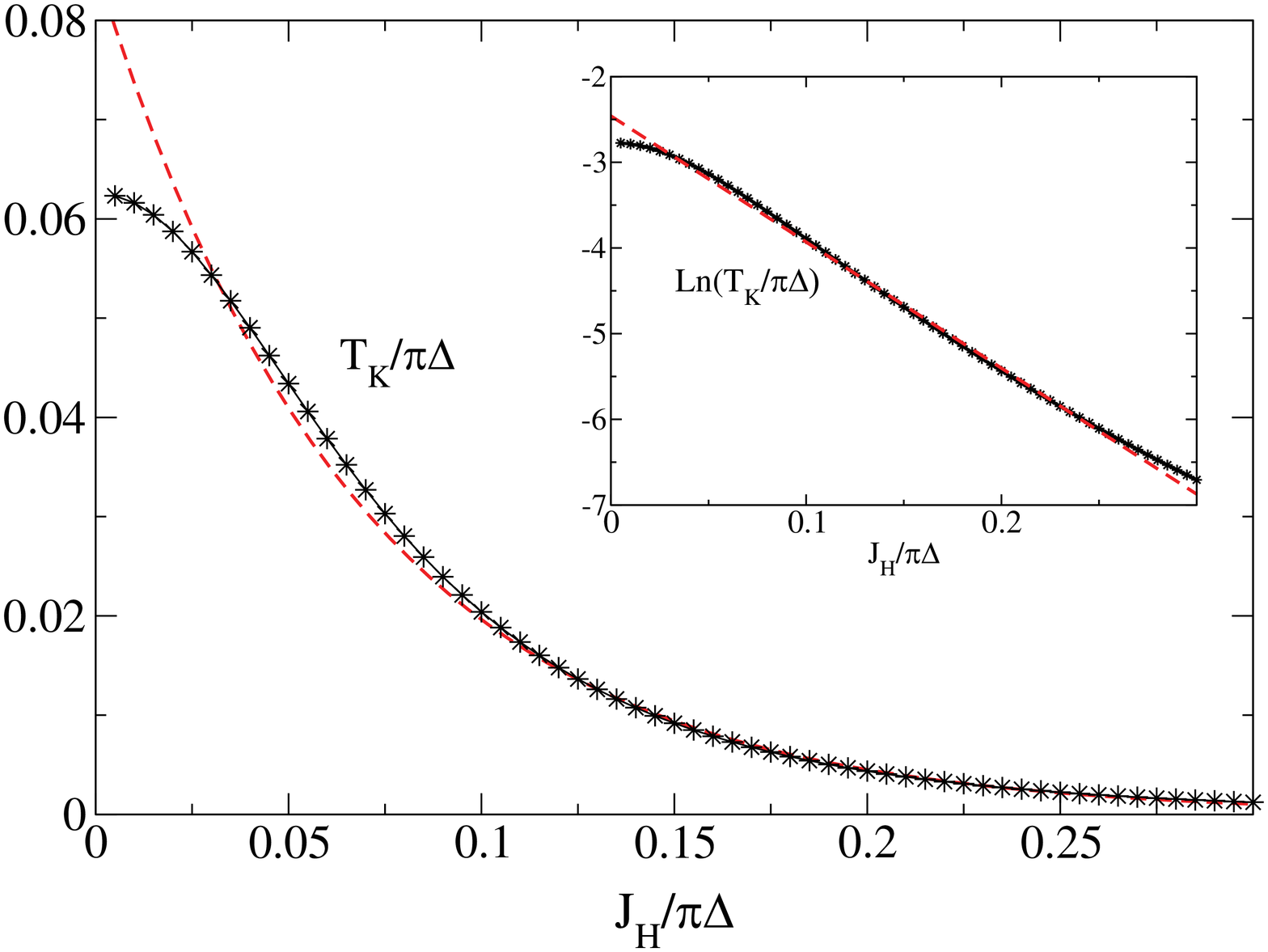}
     \caption{(Color online) A plot $T_{\rm K}/\pi\Delta$ (full curve) and $ 0.0854{\rm
         exp}(-1.49\pi^2J_{\rm H}/\pi\Delta)$ (dotted curve) versus $J_{\rm H}/\pi\Delta$
 for  $U/\pi\Delta=4$,  $\Delta=0.01$. The inset shows the  ${\rm Ln}(T_{\rm K}/\pi\Delta)$
and $-1.49\pi^2J_{\rm H}/\pi\Delta+{\rm Ln}(0.0854)$.
 } 
     \label{figtk}
   \end{center}
 \end{figure}

In the Kondo regime for  the model with $J_{\rm
  H}\ne 0$   with particle-hole symmetry we have  
$T_{\rm K}=\pi\tilde\Delta/4$. This regime occurs when $J_{\rm H}$ is large
  enough  
so that the triplet state of the impurity has a much lower energy than the other two-particle impurity
states. The effective coupling of this state to the conduction electrons, via virtual transitions
to either single particle or three particle impurity states induced by the
hybridization, leads to an exchange model of a localized spin 1
 coupled to the two channels of conduction electrons with an effective 
antiferromagnetic exchange
interaction
$J_{\rm eff}=4 V^2/(U+J_{\rm H})$. This in turn will lead to a $J_{\rm
  H}$-dependent term in the Kondo 
temperature of the form $T_{\rm K}\sim {\rm exp}(-a \pi^2J_{\rm H}/\pi\Delta)$,
where $a$ is a dimensionless numerical coefficient. This implies
that in the Kondo regime  $T_{\rm K}$ will vary exponentially with
$J_{\rm H}/\pi\Delta$. In Fig. \ref{figtk} we plot $T_{\rm K}$ from
the NRG results against $J_{\rm H}/\pi\Delta$ and compare them with an exponential
fit. The inset shows the plot of the logarithm of  $T_{\rm K}$, ${\rm Ln}(T_{\rm K}/\pi\Delta)$,
versus $J_{\rm H}/\pi\Delta$. It can be seen that the exponential form does
fit well with the results for the Kondo range $J_{\rm H}/\pi\Delta>0.1$
with the value $a=1.49$. There is a slight deviation for the largest values
of $J_{\rm H}$ shown, but the coefficient $a$ depends on the range
chosen for the  curve
 fitting.\par

\vspace*{0.7cm}
\vspace*{0.7cm}

\begin{figure}[!htbp]
   \begin{center}
     \includegraphics[width=0.4\textwidth]{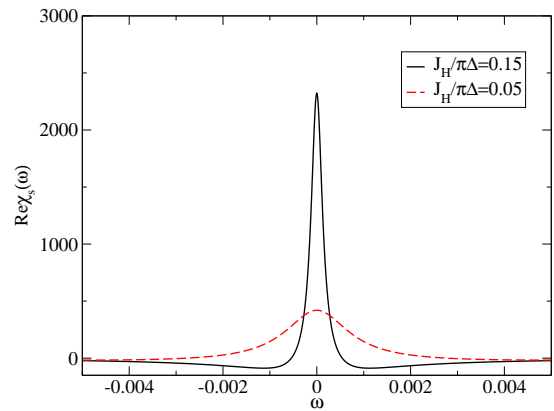}
     \caption{(Color online) A plot of the real part of the dynamic
spin susceptibility $\chi^{+-}_s(\omega)$ (in units of $8\mu_{\rm B}^2$) for   $J_{\rm H}/\pi\Delta=0.05$ and  $J_{\rm H}/\pi\Delta=0.15$
with  $U/\pi\Delta=4$,  $\Delta=0.01$.} 
     \label{fig7}
   \end{center}
 \end{figure}

\begin{figure}[!htbp]
   \begin{center}
     \includegraphics[width=0.42\textwidth]{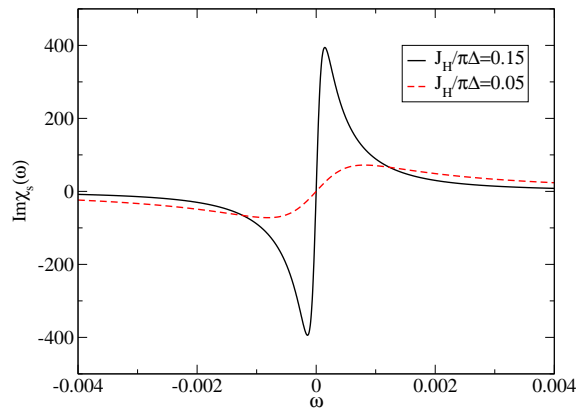}
     \caption{ (Color online) A plot of the imaginary part of the dynamic
spin susceptibility $\chi^{+-}_s(\omega)$ (in units of $8\mu_{\rm B}^2$) for   $J_{\rm H}/\pi\Delta=0.05$ and  $J_{\rm H}/\pi\Delta=0.15$
with  $U/\pi\Delta=4$,  $\Delta=0.01$. } 
     \label{fig8}
   \end{center}
 \end{figure}

Using the renormalized parameters for  values of $J_{\rm H}/\pi\Delta=0.05$
and $J_{\rm H}/\pi\Delta=0.15$  taken from fig. \ref{fig4} corresponding to
$U/\pi\Delta=4$ and $\Delta=0.01$, we have evaluated the expressions for the
dynamic
spin susceptibility given in Eq. (\ref{chiw+-}). The result for the real
  part is shown in Fig. \ref{fig7}. It illustrates the narrowing and height
  increase
of the central peak with the larger value of $J_{\rm H}$. In Fig. \ref{fig8}
the imaginary part of $\chi^{+-}_s(\omega)$ is shown. The marked increase in
the change  of the
gradient
through the origin for the larger value of $J_{\rm H}$, can be explained as a
consequence
of the Korringa-Shiba relation given in Eq. (\ref{relation3}).\par 

 \noindent

\vspace*{0.7cm}

\section{Conclusions}
The point of this study here for the n-channel Anderson model has been to show
how the renormalized
perturbation approach (RPT) can provide an asymptotically exact
way of calculating the low temperature and low frequency behavior
of the model in all parameter regimes. There have been many previous studies
of related multi-orbital impurity models using a variety 
of approaches. The general $n$-channel Anderson model with finite $U$ has not so far been solved
using the Bethe ansatz, but there are exact solutions using this technique
for the $n$-channel Kondo model coupled to a spin $S$\cite{AD84,TW84,Sch89} and the n-channel Anderson model in
the infinite $U$ limit. In the latter case the impurity occupation number
is restricted to the range $n_d\le 1$ \cite{OTW83}.  The main focus
of the work on the $n$-channel Kondo model, however,  has been on the 
over-screened case for $n\ge 2S$, where $S$ is the spin of the impurity, as
the model has a low energy non-Fermi liquid fixed point. There
have been many  NRG studies of multi-orbital models and this work has been
surveyed in the NRG review article \cite{BCP08}.
 The concern in most of
the NRG work, however,  has been with the calculation of
the one-electron spectral densities, and mainly for the models without the
Hund's rule term. 
There has also been a  recent study
for the $J_{\rm H}=0$ model using the local moment approach, which includes
NRG calculations for the case $n=2$ \cite{GGL09}, and NRG studies of
capacitively
coupled quantum dots \cite{GLK06}.
\par

 The main feature of the  RPT approach  is that  the  calculations are carried out in terms of
renormalized parameters which have a clear physical meaning in terms of the
quasiparticles and their interactions. For the $n$-channel model they  correspond to renormalizations of the parameters, $\epsilon_d$,
$\Delta$, $U$ and $J_{\rm H}$, which specify the model.
 In the strong correlation or Kondo
regime all these parameters can be determined explicitly in terms of a single
low energy parameter, the 
Kondo temperature $T_{\rm K}$. For the case $n=2$, we have been able 
to deduce the renormalized parameters from the low lying excitations in an NRG
calculation. The NRG results have confirmed the relations
we  derived
between the renormalized parameters in the Kondo regime. As we have explicit 
expressions in Eq. (\ref{shc}),  (\ref{chis}), (\ref{chiorb}) and (\ref{chic}) for the specific heat coefficient, spin, orbital and charge
susceptibilities at zero temperature, these quantities were  calculated
simply by substituting the renormalized parameters into the relevant formulae.
This procedure is  very accurate and by-passes the usual NRG method which
involves a subtraction procedure to isolate the impurity component.
As there is a large parameter space to explore we have restricted the NRG 
calculations  here to the particle-hole symmetric case. However, the RPT
results
are valid in all  parameter regimes and the behavior of the model away
from particle-hole symmetry will be the subject of a  separate publication.\par
In setting up the RPT no approximation has been made, other than the
assumption that the self-energy and its derivative are real and non-divergent
at the Fermi level $\omega=0$. This means that there is the possibility of
extending the results to higher temperatures and frequencies. Some preliminary
results have been achieved by including diagrams beyond second order
\cite{Hew01,BHO07} for the single channel model and this
topic is currently being studied. The RPT in the Keldysh formalism can also be
applied to non-equilibrium behavior and has been applied to the
calculation of the non-linear corrections to the differential conductance 
for a quantum dot \cite{Ogu05,RAH09}, including an arbitrary magnetic field
\cite{HBO05}. 
\par
 
 The RPT approach is  not restricted to impurity models,
but the  calculation of the renormalized parameters for lattice
models presents more of a problem as the NRG method cannot in general be
applied.  However, 
 for infinite dimensional lattice models
one can use the  dynamical mean field theory (DMFT) to map the model into an
effective impurity one, so the NRG method can then be used. This approach
has been used to calculate  
renormalized parameters for the one-band  Hubbard and Hubbard-Holstein models
\cite{KMH04,
BH07b}. 
 The work presented here
opens up the possibility of extending this method to the two-band Hubbard
model with a Hund's rule coupling. We have found    that the Hund's rule
term plays an important role in enhancing the magnetic response in the 2-fold
degenerate model. It is known that the single
band Hubbard model does not 
provide a basis for explaining the occurrence of ferromagnetism in 3d metals,
as it predicts a ferromagnetic ground state only in a very restricted parameter
regime, very close to half-filling and for a value of $U$ much greater than
the band width. It is likely that the inclusion of the Hund's rule coupling 
is essential to describe ferromagnetism in 3d 
materials.
\par 

\bigskip
\noindent{\bf Acknowledgment}\par
\bigskip
{We thank Akira Oguri and Johannes Bauer for helpful discussions. Two of us
  (DJGC and ACH) thank the EPRSC for support (Grant No. EP/G032181/1)
and YN thanks the JSPS Grant-in-Aid for Scientific Research (C) support  (Grant No. 20540319).
The numerical calculations were partly carried out on SX8 at  YITP in Kyoto University.
\par
\bigskip
\noindent{\bf Appendix}\par
\bigskip
For the particle-hole symmetric model with $n=2$, the model given in
Eq.
(\ref{Kondo_model}) can be derived by taking account to  order  $|V|^2$ the effects 
of
 virtual excitations from the 2-electron to the local
1-electron and 3-electron impurity states. We denote the  1-electron basis
states, $|1\uparrow\rangle$, $|1\downarrow\rangle$, $|2\uparrow\rangle$ and
$|2\downarrow\rangle$, by $|\nu\rangle$ with $\nu=1,2,3,4$ respectively. The 
2-electron
states we denote by  $|\nu,\nu'\rangle$, with $\nu\ne \nu'$ and   $|\nu',\nu\rangle$
represents the same state. This gives a six dimensional basis set. 
In terms of the Hubbard operators $X_{(\nu,\nu'):(\nu'',\nu''')}=|\nu,\nu'\rangle\langle \nu'',\nu'''|$,
the $Y_{\nu,\nu'}$ are given by 
\begin{equation}
Y_{\nu,\nu'}=\sum_{\nu''\ne \nu,\nu''\ne \nu'}(-1)^\alpha X_{(\nu,\nu''):(\nu',\nu'')},
\label{Y}
\end{equation}
for $\nu'\ge \nu$, where $\alpha=1$ if $\nu<\nu''<\nu'$, otherwise $\alpha=0$.
 The $Y_{\nu',\nu}$ for $\nu'>\nu$ can be deduced from (\ref{Y}) using
 $Y_{\nu',\nu}=(Y_{\nu,\nu'})^{\dagger}$.\par
\bigskip

\bibliography{artikel}
\bibliographystyle{h-physrev3}

\end{document}